\documentclass[pdflatex,sn-mathphys-num]{sn-jnl}

\usepackage{graphicx}%
\usepackage{multirow}%
\usepackage{amsmath,amssymb,amsfonts}%
\usepackage{amsthm}%
\usepackage{mathrsfs}%
\usepackage[title]{appendix}%
\usepackage{xcolor}%
\usepackage{textcomp}%
\usepackage{manyfoot}%
\usepackage{booktabs}%
\usepackage{algorithm}%
\usepackage{algorithmicx}%
\usepackage{algpseudocode}%
\usepackage{listings}%

\theoremstyle{thmstyleone}%
\theoremstyle{thmstyletwo}%

\theoremstyle{thmstylethree}%

\begin{document}

\title[Article Title]{Followers do not dictate the virality of news outlets on social media}


\author*[1]{\fnm{Emanuele} \sur{Sangiorgio}}\email{emanuele.sangiorgio@uniroma1.it}
\author[2]{\fnm{Matteo} \sur{Cinelli}}\email{matteo.cinelli@uniroma1.it}
\author[1,3]{\fnm{Roy} \sur{Cerqueti}}\email{roy.cerqueti@uniroma1.it}
\author[2]{\fnm{Walter} \sur{Quattrociocchi}}\email{walter.quattrociocchi@uniroma1.it}

\affil*[1]{\orgdiv{Department of Social Sciences and Economics}, \orgname{Sapienza University of Rome}, \orgaddress{\street{P.le Aldo Moro, 5}, \postcode{00185}, \state{Rome}, \country{Italy}}}

\affil[2]{\orgdiv{Department of Computer Science}, \orgname{Sapienza University of Rome}, \orgaddress{\street{ Viale Regina Elena, 295}, \postcode{00161}, \state{Rome}, \country{Italy}}}

\affil[3]{\orgdiv{GRANEM}, \orgname{Université d'Angers}, \orgaddress{\street{SFR Confluences}, \postcode{F-49000}, \state{Angers}, \country{France}}}


\abstract{Initially conceived for entertainment, social media platforms have profoundly transformed the dissemination of information and consequently reshaped the dynamics of agenda-setting.
In this scenario, understanding the factors that capture audience attention and drive viral content is crucial. Employing Gibrat's Law, which posits that an entity's growth rate is unrelated to its size, we examine the engagement growth dynamics of news outlets on social media.  Our analysis encloses the Facebook historical data of over a thousand news outlets, encompassing approximately 57 million posts in four European languages from 2008 to the end of 2022. 
We discover universal growth dynamics according to which news virality is independent of the traditional size or engagement with the outlet. Moreover, our analysis reveals a significant long-term impact of news source reliability on engagement growth, with engagement induced by unreliable sources decreasing over time. We conclude the paper by presenting a statistical model replicating the observed growth dynamics.}
\keywords{Social media $|$ Growth dynamics $|$ Attention economy }

\maketitle
\section{Introduction}  Originally designed for entertainment, social media platforms have evolved into significant channels for information dissemination \cite{kumpel2015news, walker2021news, flintham2018falling, bergstrom2018news, schmidt2017anatomy}, altering traditional agenda-setting dynamics \cite{coleman2009agenda, harder2017intermedia,feezell2018agenda, russell2014dynamics}. In this competitive landscape marked by many information sources, we aim to uncover the determinants of audience attention and the factors contributing to content virality \cite{al2019viral}, that is the propensity of content to achieve rapid diffusion and high engagement levels on social media platforms \cite{cha2010measuring,bakshy2012role, berger2012makes}.
Indeed, social media often dictate which topics become prominent while others are overlooked \cite{feezell2018agenda, barbera2019leads}. As online users tend to favor information aligning with their existing beliefs, commonly ignoring opposing viewpoints \cite{bessi2015science,zollo2017debunking, bakshy2015exposure}, this behavior can create and reinforce online 'echo chambers' \cite{del2016echo}— digital clusters of homogeneous thought where narratives are collectively shaped and solidified \cite{del2016spreading, choi2020rumor, nyhan2023like}. The magnitude of the echo chamber phenomenon and its consequent effects on polarization may vary among social media platforms \cite{cinelli2021echo}. Furthermore, many platforms implement algorithms designed to prioritize user engagement that might alter information spreading \cite{briand2021infodemics,perra2019modelling, guess2023social}, thereby exacerbating ideological divisions \cite{valensise2023drivers, gonzalez2023social, gonzalez2023asymmetric}. The rise of the attention economy is at the heart of digital discourse transformation \cite{simon1971computers, davenport2001attention, falkinger2007attention, falkinger2008limited}. In this economy, a broad spectrum of content creators, ranging from news outlets to individual influencers, vie for limited users' attention \cite{anderson2012competition, weng2012competition, tufekci2013not, lorenz2019accelerating}. Like traditional market evolution, digital stakeholders chase user engagement, converting this captured attention into tangible revenues through advertising, service offerings, and subscription models \cite{bhargava2021ethics,ciampaglia2015production}. With revenues closely linked to audience reach and engagement, understanding the growth mechanisms of digital content creators is crucial. Our research aims to unravel the dynamics of the digital ecosystem, focusing on the evolution of content consumption and audience reach. We anchor our analysis in Gibrat's Law \cite{gibrat1931inegalits}, originally formulated to explain traditional business growth, extending its application to the digital domain. The foundational premise of this law, positing that a firm's growth rate is independent of its initial size, has found relevance in various realms beyond business, like the growth patterns of city sizes \cite{rozenfeld2008laws, eeckhout2004gibrat, rose2005cities}. While various studies have explored Gibrat's Law across different contexts, yielding mixed methodologies and results \cite{mansfield1962entry, chesher1979testing, sutton1997gibrat, santarelli2006gibrat}, its implications for digital domains remain unexamined. Focusing on the supply and demand of news in the attention economy of social media platforms, we aim to determine whether the principles of proportionate growth hold in social media news dissemination. We systematically study the growth patterns of news outlets on Facebook, comparing their growth to audience sizes over different periods. For a deeper understanding of news engagement on social media, we obtain a list of news outlets from NewsGuard \cite{newsguard}, an entity recognized for tackling misinformation by assessing the credibility and reliability of news sources. After selecting all the news outlets with a Facebook account listed on NewsGuard, we use their Facebook URLs to gather their data from CrowdTangle \cite{crowdtangle}, a Facebook-owned tool that monitors interactions on public content from Facebook pages, groups, and verified profiles. This effort provides a comprehensive dataset: the Facebook historical data, from 2008 to the end of 2022, of over 1000 news outlets across four languages - English, French, German, and Italian. 
Thanks to the post-level granularity of our dataset, we can measure the growth of pages' metrics on various timescales by aggregating data according to a broader or narrower time window (daily, weekly, monthly, and quarterly), providing robust insights into online news outlets' growth dynamics. The paper is structured as follows: Initially, we define our analysis framework, investigate the growth regime, and assess its dynamics. Next, we introduce a stochastic model to replicate the observed growth patterns, illustrating the consistency between results and empirical evidence. Finally, we compare the growth of news outlets based on their information quality.
We find that the ability to create viral content and capture widespread attention is untied to the size of the information provider. Engagement follows a universal growth pattern in short-term intervals. Contrary to common belief, we observe that Followers count is not a reliable measure of a page's peaks of influence; the impact on engagement becomes apparent only over extended periods. Additionally, we discover that the unreliability of a news source negatively affects engagement growth in the long term.

\section*{Results}\label{sec_Results}
We start by defining our framework of analysis. The simplest growth model, proposed by Gibrat \cite{gibrat1931inegalits}, states that a given company's proportional growth rate is independent of its absolute initial size. His assumptions can be formalized by the following random multiplicative process for the size $S$: 
\begin{equation}
\label{Stgen}
  S_{t+\Delta t} = S_t(1+ \epsilon_t),   
\end{equation}
where $t \geq 0$ is time, $\Delta t>0$, $S_{t+\Delta t}$ and $S_t$ are the sizes at time ${t+\Delta t}$ and $t$, respectively, and $\epsilon_t$ is a random variable coming from an i.i.d. stochastic 
process uncorrelated to $S_s$ ($0 \leq s \leq t$) having mean $\mu$ and standard deviation $\sigma$. 
Due to the generic formulation of the original model, we adapt its interpretation to achieve a meaningful application in the context of social media. In terms of information spreading, virality refers to the rapid and widespread dissemination of information or content. By focusing on the extent and impact of the diffusion, virality refers to content engagement exceeding typical expectations, reaching a high level of users and interactions. In our analysis we focus on the latter facet, characterizing the growth of content performance with respect to the size of its source. The analysis is performed on two key metrics: Followers and Engagement. We first define how to assess page size and performance on social media, and our timescales of analysis. Then we evaluate the growth regime of both metrics concerning size for each timescale.

\subsection*{Metrics and Methodology for Social Media Page Analysis}
In evaluating whether the size of a page affects its growth, we first need to establish how to measure the size and its performance.
In the study of social media platforms, notably Facebook, we primarily rely on two metrics: 1) Page Followers, the number of users subscribed to a given page at the time of posting, representing a metric of reach, and 2) Engagement, encompassing the total number of users' interactions with the page's posts (that is, the sum of Likes, Comments, and Shares). The size of a page is typically inferred from its Followers count, a standard measure on such platforms. The alternative would be using Engagement, but such a choice would introduce undesired issues. Indeed, the engagement definition is inherently ambiguous: it can be a cumulative sum of interactions over the entire lifespan or a count over a specific duration, such as a week. With period-specific measures, pages' size could fluctuate too widely (spanning even across orders of magnitude), thus leading to interpretational challenges. Conversely, using a cumulative engagement count to quantify size may over-represent past performances. Consequently, we opt to use Followers as a more stable representation of page size. On the other hand, Engagement represents the page performance in terms of users' attention. Our analysis leverages varied timescales to observe growth patterns. Specifically, we consider four time-granularity: daily (D), weekly (W), monthly (M), and quarterly (Q). 
Therefore, for each page, we consider both metrics, Followers, and Engagement, according to different time windows. To measure engagement, we consider aggregated data depending on the timescale of the analysis. Focusing on the total attention received by the news outlets, we consider higher total engagement as a higher users' attention, regardless of the number of posts. Such interpretation relies on the fact that publishing more posts does not lead to more engagement if users are not interested in a topic. Likewise, getting engagement with several posts implies high attention, and using the mean value would underestimate the latter. We further validate it by showing that both measures bring equivalent results. For Followers, since they already are a cumulative value, we take only a representative data point in the time window, depending on the chosen timescale (see Materials and Methods for further details).
Transitioning between these scales offers diverse and new perspectives on growth dynamics. In the case of Followers' growth, daily measurements are deemed unsuitable due to limited variability, whereas all four scales are relevant for Engagement analysis, since news outlets are very active accounts and usually have multiple posts per day.

\subsection*{Assessing the growth regime}

\begin{figure*}[t]
       \centering
    \includegraphics[width=\textwidth]{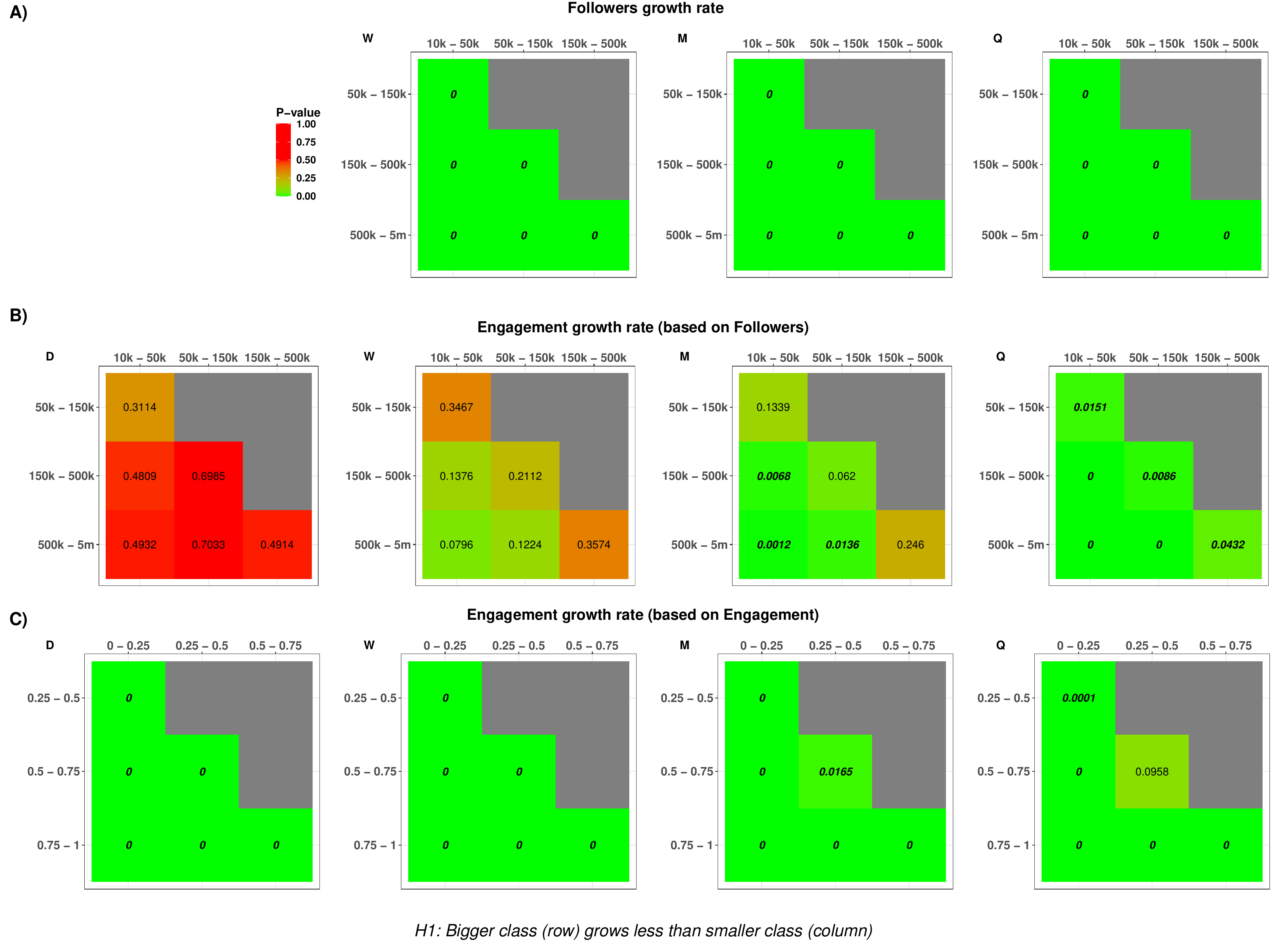}
        \caption{(A-C) p-values of Mann-Whitney U tests between classes of size for Followers and Engagement growth rate distributions. Panel titles indicate the metric being tested and the metric according to which we determine the size. 
        Row and column headers represent the class size. Bold numbers represent p-values for which we reject the hypothesis that the growth distributions do not differ, with the alternative hypothesis that the smaller class grows at a higher rate. For readability, 0 represents p-values smaller than 0.0001.}
        \label{fig:Figure_test}
\end{figure*}

\label{sec_Regime}
Based on the definition of Followers and Engagement, and according to \eqref{Stgen}, we refer to Followers and Engagement growth, respectively, as
\begin{equation}
\label{Fgib}
F_{t+\Delta t} = F_t(1+ \epsilon^{(F)}_t)
\end{equation}
\begin{equation}
\label{Egib}
E_{t+\Delta t} = E_t(1+ \epsilon^{(E)}_t),
\end{equation}
where the superscripts $(F)$ and $(E)$ point to an intuitive notation for the process $\epsilon$ with mean $ \mu_F$ and $\mu_E$ and standard deviation $\sigma_F$ and $\sigma_E$ for Followers and Engagement, respectively. Time measures the different timescales: D, W, M and Q. In \eqref{Fgib}, $F_t$ is the number of Followers at time $t$ while $E_t$ represents the number of interactions generated at time $t$. In the same way, $(1+ \epsilon^{(F)}_t)$ and $(1+ \epsilon^{(E)}_t)$ are the growth rates of $F_t$ and $E_t$, respectively. As Gibrat's Law was originally intended to explain the emergence of a log-normal distribution of sizes, we first assess that Followers and Engagement distributions comply with this assumption.  Since Followers' records on CrowdTangle start from 1/1/2018 and stop on 31/12/2022, hence relying on a five-year timespan of analysis, we take into consideration such a period for the relationship between Followers and Engagement growth. For this reason, when we consider the metrics of Followers and Engagement jointly, we restrict our analysis period to 01/01/2018 - 31/12/2022. In the analysis in which we do not account for Followers' value, we consider the entire 15-year timespan, ranging from 01/01/2008 to 31/12/2022.
See Processing Methods Section and Fig.S1 in SI Appendix for further details. Fig.S2 in SI Appendix shows distributions of both metrics at the start and end of the considered period. To assess whether growth rate distributions vary based on page size, we define four classes of pages based on their Followers, so as to have comparable populations between them over the entire period. The considered four classes of Followers are: 10K--50K, 50K--150K, 150K--500K, and 500K--5M. The bin boundaries were defined by jointly considering two aspects: a comparable number of pages between classes and actual values of Followers for which it was reasonable to account for a page as small, medium, large, or very large. As a robustness check, we reported the clustering in Fig. S3 in the SI Appendix  from which our classes and the clustering ones are predominantly overlapping. 
In evaluating growth regimes, we posit that the absence of size effects should, as an initial assumption, result in comparable growth rate distributions across different classes. To compare growth rate distributions among different classes of size, we apply, to each pair of bins, a Mann-Whitney U test on both metrics and for different timescales. Results are reported in Fig.\ref{fig:Figure_test}, while Fig.S4 in SI Appendix reports p-values of the two-tailed tests. We first inspect the case of Followers, reported in Fig.\ref{fig:Figure_test}A. Statistical tests across observed timescales consistently demonstrate that smaller pages experience greater Followers' growth than their larger counterparts. This evidence counters the notion of proportionate effect growth as described by Gibrat. In contrast, the growth dynamics of Engagement provide intriguing insights. Specifically, in Fig.\ref{fig:Figure_test}B, the growth regime is influenced by the duration of the observed timescale. In short-term observations (daily and weekly scales), engagement variation is consistent irrespective of page size. Thus, from a micro-level perspective, engagement adheres to a universal growth regime, independently of page size. However, as we transition to a monthly scale, a deviation in the regime emerges, with smaller-sized classes outperforming their larger counterparts. This deviation becomes definite on a quarterly timescale, underscoring the influence of size on long-term engagement growth.

For a comprehensive perspective, we recalibrated our Engagement analysis, reported in Fig.~\ref{fig:Figure_test}C, categorizing size based on Engagement metric. Bins are delineated by the quartiles of Engagement distribution across all pages within a specific timescale, after trimming between the 5th and the 95th percentiles. Our analysis indicates that, in this case, the system predominantly diverges from Gibrat's law of proportionate effect. Exceptions are noted for middle-sized pages (those within the 2nd and 3rd quartiles) on a quarterly timescale. Thus, in short-term observations, Engagement consistently depends on its recent performance, irrespective of page size, while the influence of Followers becomes evident with the increase of the observed timescale. We note that our results show correspondences with evidence from prior studies in different domains, such as the distribution of growth rates displaying a 'universal' form that does not depend on the size \cite{plerou1999similarities}, and the system experiencing a growth regime transition as the timescale widens \cite{qian2014origin}. These findings bear significant implications. Notably, in the short term, size does not dictate the probability of engagement growth. Extrapolating this to individual posts suggests an egalitarian landscape where every news item, irrespective of its source or the number of its Followers, has an equal propensity to go viral, that is to suddenly gain disproportionate engagement. Consequently, the mere count of Followers proves inadequate in gauging the page's potential influence. 
As a robustness check, in Fig S5 and S6 of SI Appendix, we reported two variants of the tests performed in Fig \ref{fig:Figure_test}B, showing how the results still hold by using the mean engagement value or by changing the bin boundaries. 

\subsection*{Analyzing Engagement and Followers Dynamics}
\label{sec_Dynamics}

\begin{figure*}[]
\centering
\includegraphics[width=\textwidth]{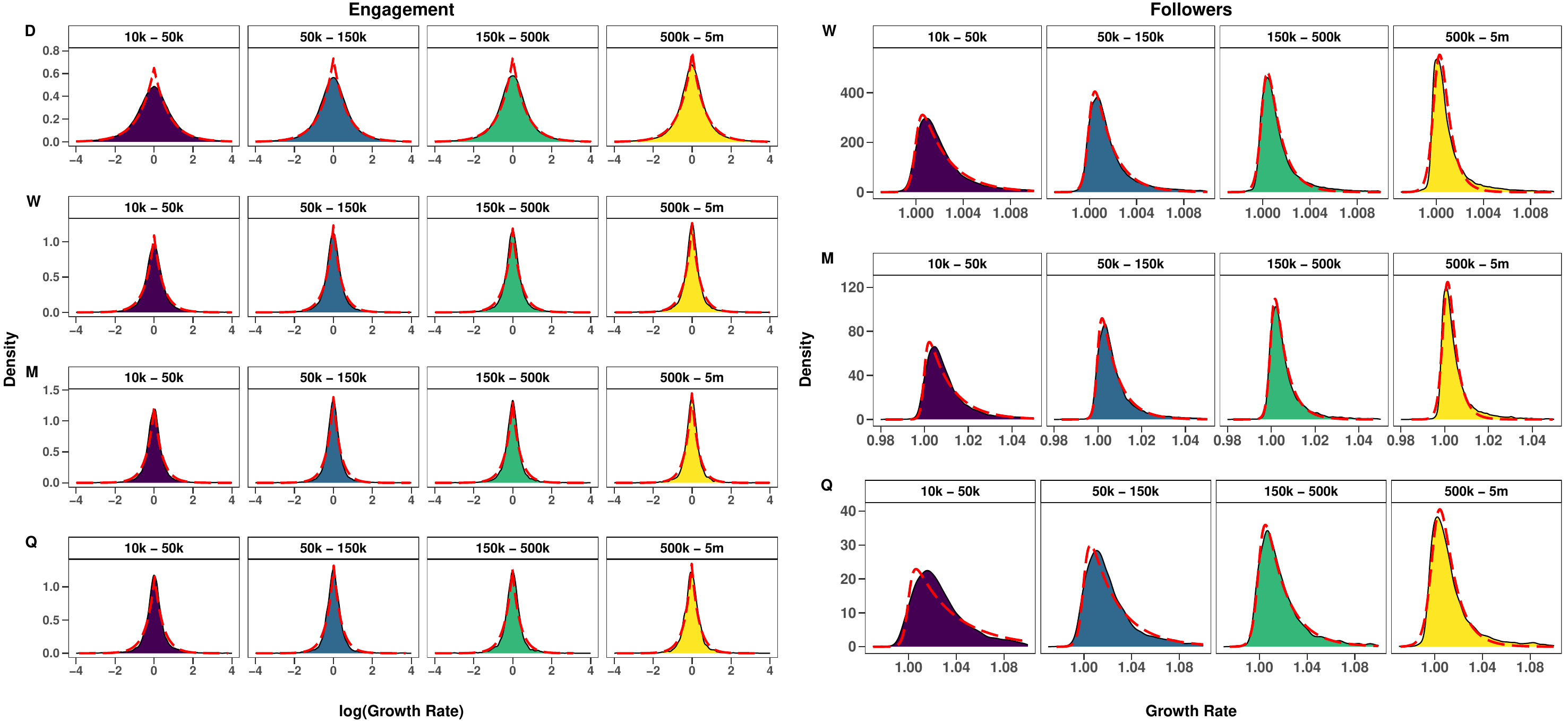}
\caption{Comparison of observed and theoretical growth rate distributions for Engagement and Followers. Red lines denote theoretical densities obtained by fitting empirical ones, with labels D, W, M, and Q indicating Daily, Weekly, Monthly, and Quarterly timescales.}
\label{fig:Figure_fit}
\end{figure*}

Empirical evidence suggests that the logarithm of many growth rate distributions often takes an exponential form. Consistently with prior studies \cite{stanley1996scaling, amaral1997scaling}, our analysis reveals that the logarithm of Engagement growth rates adheres to a particular exponential distribution known as the Laplace distribution. In contrast, the growth rates for Followers display an asymmetry, exhibiting a right-skewed distribution. Our analysis suggests a fit with a heavy-tailed distribution, specifically the Burr distribution \cite{burr1942cumulative}. Since the Burr is exclusively defined for positive values, we here employ the absolute growth rates, rather than their logarithms. A visual comparison between the observed and fitted distributions is reported in Fig.\ref{fig:Figure_fit} (see Materials and Methods section for details of the fitting procedure).
The matching of the empirical distributions of Engagement growth with the Laplace brings significant upshots. As pointed out by previous works \cite{fujiwara2003growth, fujiwara2004pareto}, growth phenomena could display a non-trivial relation between the positive and negative side of its rate distribution. The detailed balance property, or time-reversal symmetry, states that the empirical probability of changing size from one value to another is statistically the same as that for its reverse process. Statistical tests provided evidence of how the Engagement's short-term fluctuations adhere to a universal distribution, independently of the page size, with $\mu_E \to 0$ when passing from timescale Q to timescale D. Fig.S7 in SI Appendix shows parameters variation according to timescales for the considered size classes. Furthermore, the symmetry property of the Laplace distribution, with $\mu \approx 0$, directly implies the validity of detailed balance. We can draw two significant implications from this outcome. From a viewpoint of interpretation, assessing these statistical properties of short-term engagement provides a deeper understanding of news consumption dynamics, which can influence how news providers act to handle users' fluctuating attention. From a technical standpoint, ascertaining the universality of this dynamic and the probability distribution that describes it enables us to exploit it as a proxy for defining and detecting virality, namely gaining disproportionate engagement.

\subsection*{Modelling growth}\label{sec_Modelling}
Our empirical findings show that the impact of size on growth becomes evident only if observed through larger timescales and that the growth pattern is universal at the micro-level. Knowing the distributions that define the evolution of our metrics allows us to evaluate the variation of their parameters according to size and timescale. For each time scale, we can model parameters of both growth rate distributions based on Followers and Engagement values. The regression coefficients are reported in Tab~\ref{tab:table_param}. 

\begin{table}[b]
\centering
\caption{Regression coefficients and p-values of Laplace and Burr distribution parameters estimation for different timescales.}
\begin{tabular}{@{}r@{\ }lr@{\ }lr@{\ }lcc@{}}
\toprule
$\beta0$  &                   & $\beta1$ &                   & $\beta2$ &                   & Par & Time \\ \midrule
-0.109    & (0.063)           & 0.054    & (\textless 0.001) & -0.062   & (\textless 0.001) & $\mu$     & W         \\
0.073     & (0.248)           & 0.037    & (\textless 0.001) & -0.051   & (\textless 0.001) & $\mu$     & M         \\
0.384     & (\textless 0.001) & 0.031    & (\textless 0.001) & -0.065   & (\textless 0.001) & $\mu$     & Q         \\
0.613     & (\textless 0.001) & 0.027    & (0.001)           & -0.054   & (\textless 0.001) & $b$       & W         \\
0.593     & (\textless 0.001) & 0.041    & (\textless 0.001) & -0.066   & (\textless 0.001) & $b$       & M         \\
0.844     & (\textless 0.001) & 0.056    & (\textless 0.001) & -0.094   & (\textless 0.001) & $b$       & Q         \\
8420.469 & (\textless 0.001) & -372.77  & (0.025)           & -        &                   & $c$       & W         \\
2550.01   & (\textless 0.001) & -127.559 & (0.014)           & -        &                   & $c$       & M         \\
1053.905 & (0.002)           & -56.113  & (0.017)           & -        &                   & $c$       & Q         \\
-0.778    & (\textless 0.001) & 0.083    & (\textless 0.001) & -        &                   & $k$       & W         \\
-0.751    & (\textless 0.001) & 0.078    & (\textless 0.001) & -        &                   & $k$       & M         \\
-0.714    & (0.001)           & 0.073    & (\textless 0.001) & -        &                   & $k$       & Q         \\ \bottomrule
\end{tabular}
\label{tab:table_param}
\end{table}

We can thereby simulate growth on the chosen timescale, given two starting values of Followers and Engagement, $F_{0}$ and $E_{0}$, by iteratively sampling growth rates value from the distributions modeled using the parameters specified in Case 1 and Case 2.

\subsubsection*{Case 1: Engagement growth}
We consider the dynamics described in \eqref{Egib} with $\epsilon^{(E)} $ following the Laplace distribution in \eqref{Laplace} (see Materials and Methods), with:
\begin{equation}
\label{eq_u}
    \mu = \beta0_{\mu} + \beta1_{\mu} ln(F_{t}) + \beta2_{\mu} ln(E_{t})
\end{equation}
\begin{equation}
\label{eq_b}
    b = \beta0_{b}  + \beta1_{b} ln(F_{t}) + \beta2_{b} ln(E_{t})
\end{equation}

\subsubsection*{Case 2: Followers growth}
We consider \eqref{Fgib}, being $\epsilon^{(F)} $ the logarithm of the growth rate behaving according to the Burr distribution in \eqref{Burr} (see Materials and Methods), with:
\begin{equation}\label{eq_c}
    c = \beta0_{c} + \beta1_{c} ln(F_{t})
\end{equation}
\begin{equation}
\label{eq_k}
    k = \beta0_{k} + \beta1_{k} ln(F_{t})
\end{equation}

\begin{figure*}[]
\centering
\includegraphics[width=0.7\textwidth]{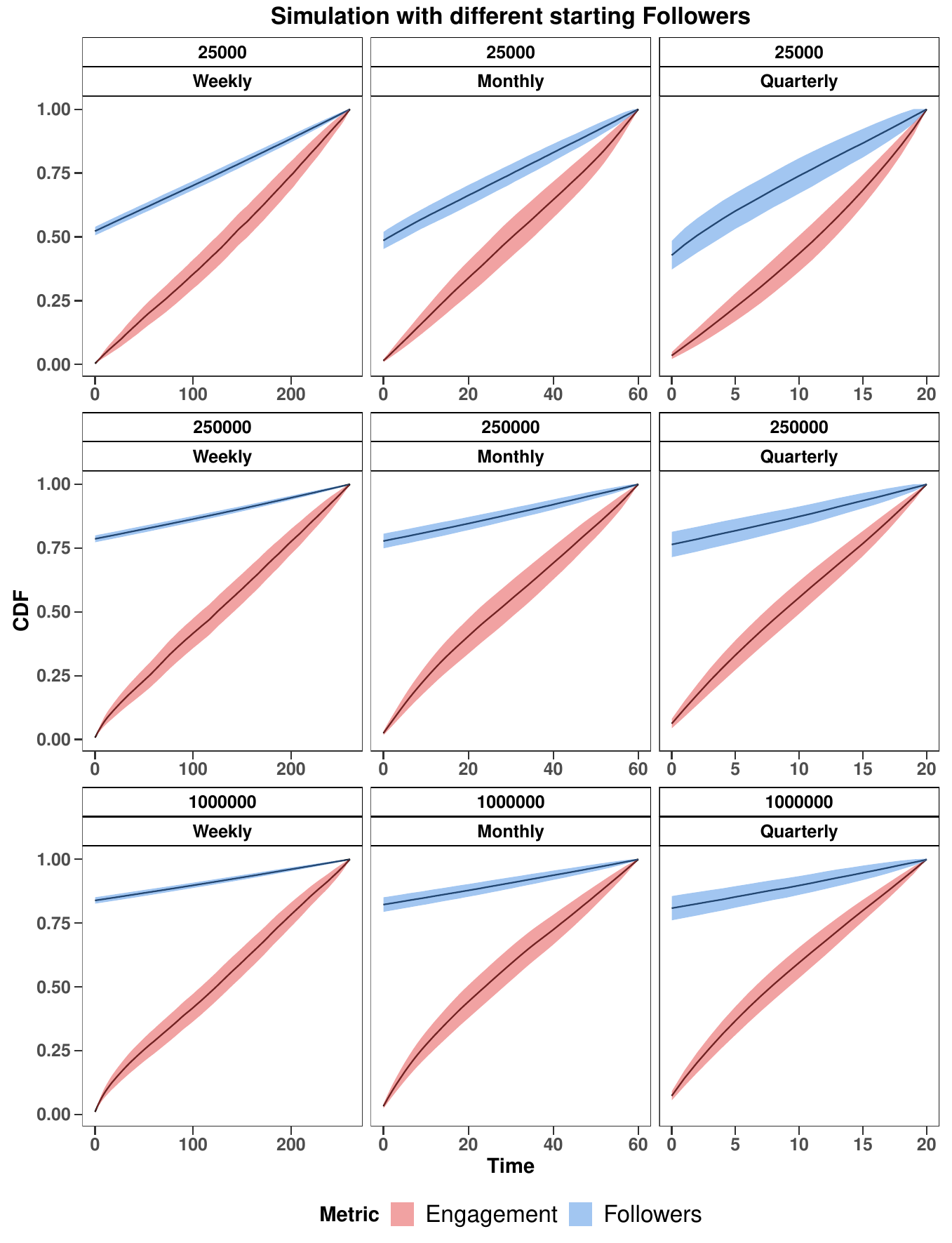}
\caption{Results of growth simulation with different starting sizes. Sub-plot headers indicate the Followers starting value and the related timescale. Solid lines represent the mean cumulative distribution function value of the iteration time, shades represent the corresponding standard error.}
\label{fig:Figure_sim}
\end{figure*}

Results of simulations are shown in Fig. \ref{fig:Figure_sim}, representing the evolution of both metrics for different starting sizes (Followers) on three timescales used for the analysis (W, M, and Q). We selected three starting sizes representing pages with low, medium, and high number of Followers, i.e., 25K, 250K, and 1M, respectively. As results show, by observing the system on a weekly timescale, the engagement shows a basically steady evolution for all three sizes. As we extend the observed timescale, by passing from weekly to quarterly, the engagement growth of small pages begins to exhibit convex behavior, while the growth curve of big pages shifts toward concavity, providing evidence of how Followers impact the engagement evolution only over long-term intervals. On the other hand, Followers of smaller pages always grow faster than bigger ones in each timescale. Our simulations consistently match empirical evidences. Despite the model's engagement growth probability being based on Followers, the universal characteristic of the process at the temporal micro-scale level is evident. These results highlight the limitation of using Followers as the sole metric to gauge overall page influence.  Short-term outcomes seem to derive from a uniform stochastic process, possibly elucidating the influence of algorithms on user news consumption behaviors. While this suggests an environment where all content providers might be on an equal stand regarding visibility, it also necessitates continuous monitoring to mitigate the spread of harmful content, such as misinformation. Basing influence assessments solely on the number of Followers can lead to oversight. The potential presence of 'one-time' or 'hidden influencers' — entities with a disproportionate influence relative to their Follower count — needs attention. The missing of a clear engagement effect on Follower growth, the lack of significance of ${\beta2}$ on Burr's parameters variation, further emphasizes this, indicating that heightened interactions do not necessarily translate to a corresponding increase in Followers or sustained reach.

\subsection*{Growth and Information Quality}

\begin{figure*}[b]
       \centering
       \includegraphics[width=0.9\textwidth]{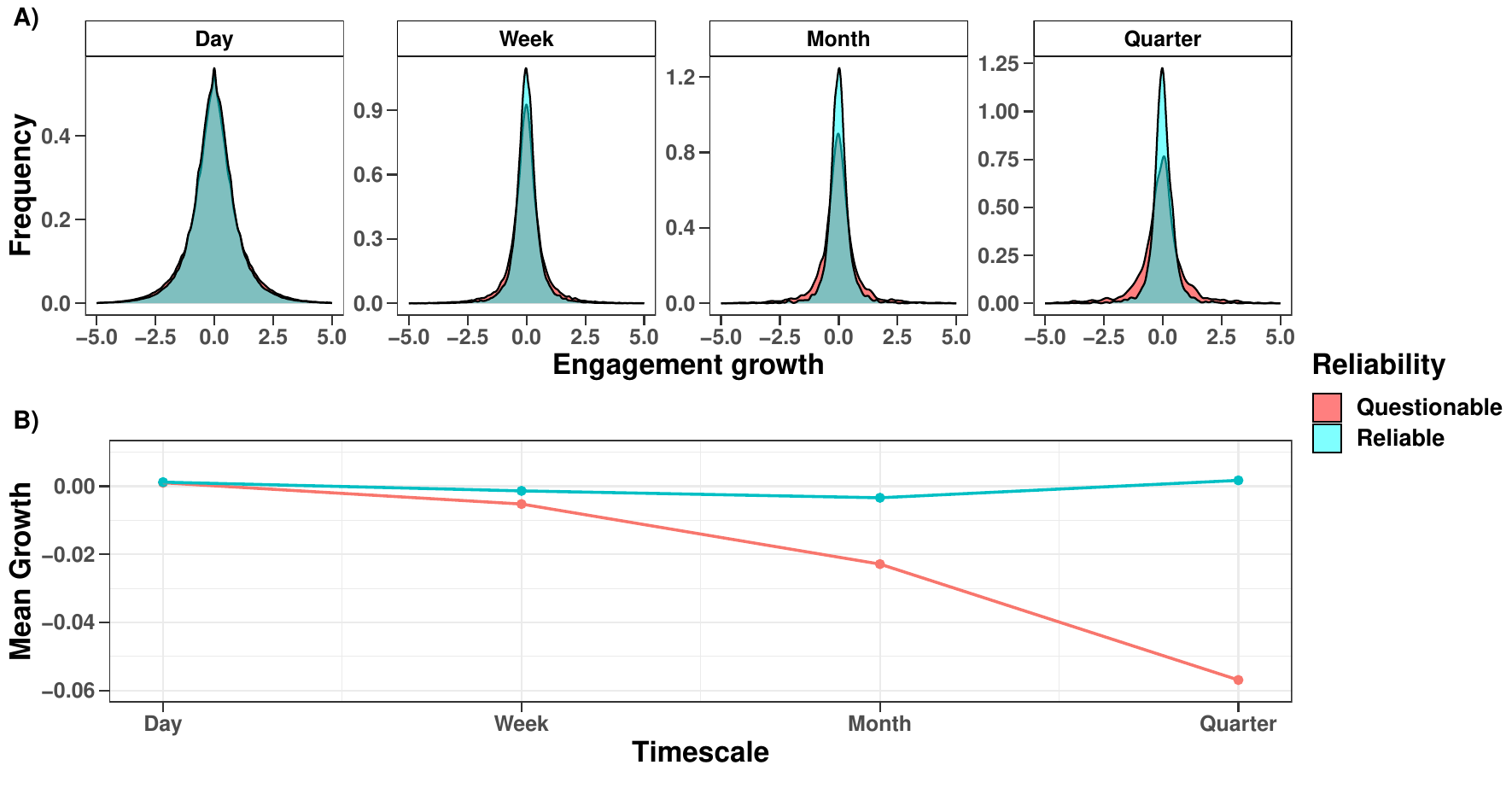}
       \caption{A) Comparison of Engagement growth rate distributions of Questionable and Reliable pages for different timescales. B) Mean growths of Questionable and Reliable pages across increasing timescales.}
        \label{fig:Figure_reliability}
\end{figure*} 

\label{sec_Reliability}
The effects of external factors on engagement growth manifest in the long term. Potential explanations for differences in page growth could be plenty, though one of the most relevant for society is the propensity of news outlets to produce unverified news and misinformation. For this reason, we conclude our analysis by comparing two sub-samples of pages representing reliable and questionable ones. Since we do not account for Followers value, this analysis encompasses the entire pages' lifespan. The classification is performed based on reliability scores provided by Newsguard. Since our dataset comprises 898 reliable sources and 131 non-reliable ones, we performed a sampling of 131 reliable sources to obtain two comparable samples. See Materials and Methods and Fig.S8 in SI Appendix for further details about the reliability ratings and the sampling procedure. Fig.\ref{fig:Figure_reliability} shows growth rate distributions of engagement and their evolution across the various timescales. Tab.S1 in SI Appendix shows the p-values of Mann-Whitney U tests between the two sub-samples, as in our previous analyses. Both graphic representations and tests display how the trustworthiness of the news source plays a crucial role, as the engagement of unreliable pages progressively decreases as the time scale widens. Anew, the short-term fluctuations follow a universal dynamic, and neither the reliability turns out to determine growth differences. The long-term divergence, which may result from both users' behavior and platform moderation policies, along with the inherent randomness of short-term fluctuations, highlight the importance of continuous efforts to monitor the consumption of sensitive content, such as misinformation.

\subsection*{Conclusions}
In historical media landscapes, prominent news outlets predominantly influenced agenda-setting, their reach determining the flow and focus of public discourse. However, the emergence of social media platforms—designed more for entertainment than information spreading—has reshaped this dynamic. 
While many assume larger outlets and their inherent reach would dominate social media discourse, our research challenges this perspective. 
Analyzing engagement metrics across diverse news outlets on Facebook, we find that news virality, namely a disproportionate growth of engagement, is not strictly tied to the traditional size of the outlet. Instead, a myriad of factors may drive online discourse: rapid user engagement \cite{etta2023characterizing}, the reinforcing nature of echo chambers \cite{cinelli2021echo}, the amplifying power of influencers \cite{bakshy2011everyone,bakshy2012role}, the emotional resonance of content \cite{del2016echo}, and even artificial amplification via bots \cite{cresci2020decade}. 
This complex web of drivers, some of which exhibit random behaviors, defies conventional models of media influence. Indeed, understanding the dynamics of the attention economy is pivotal for charting the trajectory of content creators on platforms like Facebook. 
In this work, we analyze a massive dataset composed of 57 million posts comprising the entire Facebook history, spanning 15 years, of over 1000 news outlets. 
In particular, this study took a deep dive into these dynamics, evaluating the applicability of Gibrat's Law — a principle traditionally applied to business growth — in social media content creation.
Empirical results provided a nuanced understanding of growth patterns. We observe that the likelihood of generating viral content and capturing widespread attention is independent of the information provider's size. Indeed, engagement adheres to a universal growth pattern in short-term intervals. This pattern shifts as the analysis extends to longer timescales like monthly and quarterly intervals, where size effects begin to manifest. We validated this dynamic by comparing news outlets’ growth based on their information quality, providing evidence on how, though the unreliability of the news source negatively impacts engagement growth, its effect only manifests in the long term. Another significant observation challenges conventional wisdom: Followers count is not a sufficient indicator of a page's potential influence, and its actual impact only emerges over extended periods. 
Our examination of growth dynamics further elucidated these insights. After detecting their probability distributions, we evaluated their behavior according to size and timescale. We developed a stochastic model validating our empirical findings, emphasizing that Followers do not always depict actual influence or engagement potential in the short term.
This brings broader implications in the context of agenda-setting dynamics in the social media era. Our study shows that contrary to traditional media, influence is not strictly tied to size or following in the digital realm. 
This stochastic nature of short-term engagement suggests an environment where all content, irrespective of its source, stands a roughly equal chance of capturing attention, possibly elucidating the influence of algorithms on users' news consumption.
This democratization of potential attention influences how narratives and agendas are set, with even smaller entities having the power to shape discourse. However, it also emphasizes the importance of vigilant monitoring mechanisms, given the risk of rapid misinformation or harmful content spread. 
Our research highlights the intricate dynamics of growth in the digital attention economy, revealing how traditional metrics may not align with real-world influence. It also offers key insights into how the modern agenda-setting dynamics are being reshaped in the era of social media. These findings are precious for content creators, platform designers, and policymakers as they navigate the complexities of the digital age.

\section{Materials and Methods}

\subsection{Selecting Followers' value} In determining the Followers' value for each time window, we selected the closest observation to a given time point of the window, which we referred to as a 'representative point in time' since it varies depending on the timescale. For the weekly scale, we selected the value on the minimum observed date of the week. For the monthly scale, we selected the value of the closest observed date to the central point of the month (the 15th day). For the quarterly scale, we selected the value of the farther observed date.

\subsection*{Labeling of Media Sources}
The reliability labeling of news outlets is based on the trust ratings provided by Newsguard \cite{newsguard}. Each site is rated using nine basic, apolitical criteria of journalistic practice, related to credibility and transparency. Based on the nine criteria, each site gets a trust score of 0-100 points. NewsGuard labels the source as Trustable if the resulting score equals or exceeds 60. The total number of news outlets for which we have a trust rating is 1029.

\subsection*{Parameters Estimation}
\label{matmed_Fit}
Here we provide details about the fitting procedure of distributions reported in Fig \ref{fig:Figure_fit} of sections Analyzing Engagement and Followers Dynamics, and Modelling Growth.

\subsubsection*{Laplace Distribution}
The probability density function of the Engagement growth rate is the Laplace distribution, expressed as:
\begin{equation}
\label{Laplace}
f_L(x | \mu , b) = \frac{1}{2b} \exp\left(-\frac{|x-\mu|}{b}\right) ,
\end{equation}
where $x \in \mathbb{R}$ and $\mu$ and $b$ are parameters to be calibrated. In this respect, the parameters of the Laplace distribution can be derived analytically from the mean $\mu_X$ and standard deviation $\sigma_X$ of the empirical distribution $X$, since 

\begin{align}
    \mu = \mu_X;\,   b = \frac{\sigma_X}{\sqrt{2}}
\end{align}

\subsubsection*{Burr Distribution}
The probability density function for Followers' growth is described by the Burr distribution, whose density is:
\begin{equation}
\label{Burr}
f_B(x | c, k) = ck\frac{x^{c-1}}{(1+x^{c})^{k+1}},
\end{equation}
where $x \in \mathbb{R}$ and $c$ and $k$ are scalars to be calibrated.
In particular, such parameters are evaluated by fitting the empirical cumulative distribution function of the observed growth rates with the Burr's one.

\subsubsection*{Regression of distribution parameters} 
To model parameters variation according to Followers and Engagement values, we first applied the fitting procedure described above to the growth distributions of the sub-samples obtained by binning based on Followers and Engagement, after trimming within the 5th and 95th percentiles of our observed distributions, in each timescale. After obtaining the parameters of the Laplace and Burr distribution of each sub-sample, we performed the parameter regression as described by equations [\ref{eq_u}], [\ref{eq_b}], [\ref{eq_c}], and [\ref{eq_k}], of section Modelling growth. 

\subsection*{Sampling of reliable news outlets} According to NewsGuard ratings, our dataset comprises 898 reliable sources and 131 non-reliable ones. We performed a sampling of 131 reliable sources to have two comparable samples. To obtain similar structural characteristics that are not being tested, namely Followers and Page's lifespan, the distance is computed using the maximum number of Followers and the page's creation date as distance variables, since most pages' last observations coincide with the end of the analyzed period. The resulting sample is obtained by computing the Euclidean distances of all the possible couples of Questionable and Reliable pages in a two-dimensional space, using Followers and Lifespan as space variables. Then we selected the partition of 131 Reliable pages for which the sum of their Euclidean distances from the 131 Questionable pages was minimized. 

\section{Author contributions statement}
E.S. and M.C. designed the paper; E.S. performed data collection and analysis; M.C., R.C. and W.Q. supervised the project; All authors wrote the paper.

\section{Data availability}
The data collection and analysis process are compliant with the terms and conditions imposed by Crowdtangle \cite{crowdtangle}. Therefore, the results described in this paper cannot be exploited to infer the identity of the accounts involved. CrowdTangle does not include paid ads unless those ads began as organic, non-paid posts that were subsequently “boosted” using Facebook’s advertising tools. It also does not include activity on private accounts, or posts made visible only to specific groups of followers. 

\bibliography{bibliography}


\begin{thebibliography}{58}
\ifx \bisbn   \undefined \def \bisbn  #1{ISBN #1}\fi
\ifx \binits  \undefined \def \binits#1{#1}\fi
\ifx \bauthor  \undefined \def \bauthor#1{#1}\fi
\ifx \batitle  \undefined \def \batitle#1{#1}\fi
\ifx \bjtitle  \undefined \def \bjtitle#1{#1}\fi
\ifx \bvolume  \undefined \def \bvolume#1{\textbf{#1}}\fi
\ifx \byear  \undefined \def \byear#1{#1}\fi
\ifx \bissue  \undefined \def \bissue#1{#1}\fi
\ifx \bfpage  \undefined \def \bfpage#1{#1}\fi
\ifx \blpage  \undefined \def \blpage #1{#1}\fi
\ifx \burl  \undefined \def \burl#1{\textsf{#1}}\fi
\ifx \doiurl  \undefined \def \doiurl#1{\url{https://doi.org/#1}}\fi
\ifx \betal  \undefined \def \betal{\textit{et al.}}\fi
\ifx \binstitute  \undefined \def \binstitute#1{#1}\fi
\ifx \binstitutionaled  \undefined \def \binstitutionaled#1{#1}\fi
\ifx \bctitle  \undefined \def \bctitle#1{#1}\fi
\ifx \beditor  \undefined \def \beditor#1{#1}\fi
\ifx \bpublisher  \undefined \def \bpublisher#1{#1}\fi
\ifx \bbtitle  \undefined \def \bbtitle#1{#1}\fi
\ifx \bedition  \undefined \def \bedition#1{#1}\fi
\ifx \bseriesno  \undefined \def \bseriesno#1{#1}\fi
\ifx \blocation  \undefined \def \blocation#1{#1}\fi
\ifx \bsertitle  \undefined \def \bsertitle#1{#1}\fi
\ifx \bsnm \undefined \def \bsnm#1{#1}\fi
\ifx \bsuffix \undefined \def \bsuffix#1{#1}\fi
\ifx \bparticle \undefined \def \bparticle#1{#1}\fi
\ifx \barticle \undefined \def \barticle#1{#1}\fi
\bibcommenthead
\ifx \bconfdate \undefined \def \bconfdate #1{#1}\fi
\ifx \botherref \undefined \def \botherref #1{#1}\fi
\ifx \url \undefined \def \url#1{\textsf{#1}}\fi
\ifx \bchapter \undefined \def \bchapter#1{#1}\fi
\ifx \bbook \undefined \def \bbook#1{#1}\fi
\ifx \bcomment \undefined \def \bcomment#1{#1}\fi
\ifx \oauthor \undefined \def \oauthor#1{#1}\fi
\ifx \citeauthoryear \undefined \def \citeauthoryear#1{#1}\fi
\ifx \endbibitem  \undefined \def \endbibitem {}\fi
\ifx \bconflocation  \undefined \def \bconflocation#1{#1}\fi
\ifx \arxivurl  \undefined \def \arxivurl#1{\textsf{#1}}\fi
\csname PreBibitemsHook\endcsname

\bibitem[\protect\citeauthoryear{K{\"u}mpel et~al.}{2015}]{kumpel2015news}
\begin{barticle}
\bauthor{\bsnm{K{\"u}mpel}, \binits{A.S.}},
\bauthor{\bsnm{Karnowski}, \binits{V.}},
\bauthor{\bsnm{Keyling}, \binits{T.}}:
\batitle{News sharing in social media: A review of current research on news sharing users, content, and networks}.
\bjtitle{Social media+ society}
\bvolume{1}(\bissue{2}),
\bfpage{2056305115610141}
(\byear{2015})
\end{barticle}
\endbibitem

\bibitem[\protect\citeauthoryear{Walker and Matsa}{2021}]{walker2021news}
\begin{botherref}
\oauthor{\bsnm{Walker}, \binits{M.}},
\oauthor{\bsnm{Matsa}, \binits{K.E.}}:
News consumption across social media in 2021.
Technical report,
Pew Research Center
(2021)
\end{botherref}
\endbibitem

\bibitem[\protect\citeauthoryear{Flintham et~al.}{2018}]{flintham2018falling}
\begin{bchapter}
\bauthor{\bsnm{Flintham}, \binits{M.}},
\bauthor{\bsnm{Karner}, \binits{C.}},
\bauthor{\bsnm{Bachour}, \binits{K.}},
\bauthor{\bsnm{Creswick}, \binits{H.}},
\bauthor{\bsnm{Gupta}, \binits{N.}},
\bauthor{\bsnm{Moran}, \binits{S.}}:
\bctitle{Falling for fake news: investigating the consumption of news via social media}.
In: \bbtitle{Proceedings of the 2018 CHI Conference on Human Factors in Computing Systems},
pp. \bfpage{1}--\blpage{10}
(\byear{2018})
\end{bchapter}
\endbibitem

\bibitem[\protect\citeauthoryear{Bergstr{\"o}m and Jervelycke~Belfrage}{2018}]{bergstrom2018news}
\begin{barticle}
\bauthor{\bsnm{Bergstr{\"o}m}, \binits{A.}},
\bauthor{\bsnm{Jervelycke~Belfrage}, \binits{M.}}:
\batitle{News in social media: Incidental consumption and the role of opinion leaders}.
\bjtitle{Digital journalism}
\bvolume{6}(\bissue{5}),
\bfpage{583}--\blpage{598}
(\byear{2018})
\end{barticle}
\endbibitem

\bibitem[\protect\citeauthoryear{Schmidt et~al.}{2017}]{schmidt2017anatomy}
\begin{barticle}
\bauthor{\bsnm{Schmidt}, \binits{A.L.}},
\bauthor{\bsnm{Zollo}, \binits{F.}},
\bauthor{\bsnm{Del~Vicario}, \binits{M.}},
\bauthor{\bsnm{Bessi}, \binits{A.}},
\bauthor{\bsnm{Scala}, \binits{A.}},
\bauthor{\bsnm{Caldarelli}, \binits{G.}},
\bauthor{\bsnm{Stanley}, \binits{H.E.}},
\bauthor{\bsnm{Quattrociocchi}, \binits{W.}}:
\batitle{Anatomy of news consumption on facebook}.
\bjtitle{Proceedings of the National Academy of Sciences}
\bvolume{114}(\bissue{12}),
\bfpage{3035}--\blpage{3039}
(\byear{2017})
\end{barticle}
\endbibitem

\bibitem[\protect\citeauthoryear{Coleman et~al.}{2009}]{coleman2009agenda}
\begin{bchapter}
\bauthor{\bsnm{Coleman}, \binits{R.}},
\bauthor{\bsnm{McCombs}, \binits{M.}},
\bauthor{\bsnm{Shaw}, \binits{D.}},
\bauthor{\bsnm{Weaver}, \binits{D.}}:
\bctitle{Agenda setting}.
In: \bbtitle{The Handbook of Journalism Studies},
pp. \bfpage{167}--\blpage{180}.
\bpublisher{Routledge}, \blocation{???}
(\byear{2009})
\end{bchapter}
\endbibitem

\bibitem[\protect\citeauthoryear{Harder et~al.}{2017}]{harder2017intermedia}
\begin{barticle}
\bauthor{\bsnm{Harder}, \binits{R.A.}},
\bauthor{\bsnm{Sevenans}, \binits{J.}},
\bauthor{\bsnm{Van~Aelst}, \binits{P.}}:
\batitle{Intermedia agenda setting in the social media age: How traditional players dominate the news agenda in election times}.
\bjtitle{The international journal of press/politics}
\bvolume{22}(\bissue{3}),
\bfpage{275}--\blpage{293}
(\byear{2017})
\end{barticle}
\endbibitem

\bibitem[\protect\citeauthoryear{Feezell}{2018}]{feezell2018agenda}
\begin{barticle}
\bauthor{\bsnm{Feezell}, \binits{J.T.}}:
\batitle{Agenda setting through social media: The importance of incidental news exposure and social filtering in the digital era}.
\bjtitle{Political Research Quarterly}
\bvolume{71}(\bissue{2}),
\bfpage{482}--\blpage{494}
(\byear{2018})
\end{barticle}
\endbibitem

\bibitem[\protect\citeauthoryear{Russell~Neuman et~al.}{2014}]{russell2014dynamics}
\begin{barticle}
\bauthor{\bsnm{Russell~Neuman}, \binits{W.}},
\bauthor{\bsnm{Guggenheim}, \binits{L.}},
\bauthor{\bsnm{Mo~Jang}, \binits{S.a.}},
\bauthor{\bsnm{Bae}, \binits{S.Y.}}:
\batitle{The dynamics of public attention: Agenda-setting theory meets big data}.
\bjtitle{Journal of communication}
\bvolume{64}(\bissue{2}),
\bfpage{193}--\blpage{214}
(\byear{2014})
\end{barticle}
\endbibitem

\bibitem[\protect\citeauthoryear{Al-Rawi}{2019}]{al2019viral}
\begin{barticle}
\bauthor{\bsnm{Al-Rawi}, \binits{A.}}:
\batitle{Viral news on social media}.
\bjtitle{Digital journalism}
\bvolume{7}(\bissue{1}),
\bfpage{63}--\blpage{79}
(\byear{2019})
\end{barticle}
\endbibitem

\bibitem[\protect\citeauthoryear{Cha et~al.}{2010}]{cha2010measuring}
\begin{bchapter}
\bauthor{\bsnm{Cha}, \binits{M.}},
\bauthor{\bsnm{Haddadi}, \binits{H.}},
\bauthor{\bsnm{Benevenuto}, \binits{F.}},
\bauthor{\bsnm{Gummadi}, \binits{K.}}:
\bctitle{Measuring user influence in twitter: The million follower fallacy}.
In: \bbtitle{Proceedings of the International AAAI Conference on Web and Social Media},
vol. \bseriesno{4},
pp. \bfpage{10}--\blpage{17}
(\byear{2010})
\end{bchapter}
\endbibitem

\bibitem[\protect\citeauthoryear{Bakshy et~al.}{2012}]{bakshy2012role}
\begin{bchapter}
\bauthor{\bsnm{Bakshy}, \binits{E.}},
\bauthor{\bsnm{Rosenn}, \binits{I.}},
\bauthor{\bsnm{Marlow}, \binits{C.}},
\bauthor{\bsnm{Adamic}, \binits{L.}}:
\bctitle{The role of social networks in information diffusion}.
In: \bbtitle{Proceedings of the 21st International Conference on World Wide Web},
pp. \bfpage{519}--\blpage{528}
(\byear{2012})
\end{bchapter}
\endbibitem

\bibitem[\protect\citeauthoryear{Berger and Milkman}{2012}]{berger2012makes}
\begin{barticle}
\bauthor{\bsnm{Berger}, \binits{J.}},
\bauthor{\bsnm{Milkman}, \binits{K.L.}}:
\batitle{What makes online content viral?}
\bjtitle{Journal of marketing research}
\bvolume{49}(\bissue{2}),
\bfpage{192}--\blpage{205}
(\byear{2012})
\end{barticle}
\endbibitem

\bibitem[\protect\citeauthoryear{Barber{\'a} et~al.}{2019}]{barbera2019leads}
\begin{barticle}
\bauthor{\bsnm{Barber{\'a}}, \binits{P.}},
\bauthor{\bsnm{Casas}, \binits{A.}},
\bauthor{\bsnm{Nagler}, \binits{J.}},
\bauthor{\bsnm{Egan}, \binits{P.J.}},
\bauthor{\bsnm{Bonneau}, \binits{R.}},
\bauthor{\bsnm{Jost}, \binits{J.T.}},
\bauthor{\bsnm{Tucker}, \binits{J.A.}}:
\batitle{Who leads? who follows? measuring issue attention and agenda setting by legislators and the mass public using social media data}.
\bjtitle{American Political Science Review}
\bvolume{113}(\bissue{4}),
\bfpage{883}--\blpage{901}
(\byear{2019})
\end{barticle}
\endbibitem

\bibitem[\protect\citeauthoryear{Bessi et~al.}{2015}]{bessi2015science}
\begin{barticle}
\bauthor{\bsnm{Bessi}, \binits{A.}},
\bauthor{\bsnm{Coletto}, \binits{M.}},
\bauthor{\bsnm{Davidescu}, \binits{G.A.}},
\bauthor{\bsnm{Scala}, \binits{A.}},
\bauthor{\bsnm{Caldarelli}, \binits{G.}},
\bauthor{\bsnm{Quattrociocchi}, \binits{W.}}:
\batitle{Science vs conspiracy: Collective narratives in the age of misinformation}.
\bjtitle{PloS one}
\bvolume{10}(\bissue{2}),
\bfpage{0118093}
(\byear{2015})
\end{barticle}
\endbibitem

\bibitem[\protect\citeauthoryear{Zollo et~al.}{2017}]{zollo2017debunking}
\begin{barticle}
\bauthor{\bsnm{Zollo}, \binits{F.}},
\bauthor{\bsnm{Bessi}, \binits{A.}},
\bauthor{\bsnm{Del~Vicario}, \binits{M.}},
\bauthor{\bsnm{Scala}, \binits{A.}},
\bauthor{\bsnm{Caldarelli}, \binits{G.}},
\bauthor{\bsnm{Shekhtman}, \binits{L.}},
\bauthor{\bsnm{Havlin}, \binits{S.}},
\bauthor{\bsnm{Quattrociocchi}, \binits{W.}}:
\batitle{Debunking in a world of tribes}.
\bjtitle{PloS one}
\bvolume{12}(\bissue{7}),
\bfpage{0181821}
(\byear{2017})
\end{barticle}
\endbibitem

\bibitem[\protect\citeauthoryear{Bakshy et~al.}{2015}]{bakshy2015exposure}
\begin{barticle}
\bauthor{\bsnm{Bakshy}, \binits{E.}},
\bauthor{\bsnm{Messing}, \binits{S.}},
\bauthor{\bsnm{Adamic}, \binits{L.A.}}:
\batitle{Exposure to ideologically diverse news and opinion on facebook}.
\bjtitle{Science}
\bvolume{348}(\bissue{6239}),
\bfpage{1130}--\blpage{1132}
(\byear{2015})
\end{barticle}
\endbibitem

\bibitem[\protect\citeauthoryear{Del~Vicario et~al.}{2016a}]{del2016echo}
\begin{barticle}
\bauthor{\bsnm{Del~Vicario}, \binits{M.}},
\bauthor{\bsnm{Vivaldo}, \binits{G.}},
\bauthor{\bsnm{Bessi}, \binits{A.}},
\bauthor{\bsnm{Zollo}, \binits{F.}},
\bauthor{\bsnm{Scala}, \binits{A.}},
\bauthor{\bsnm{Caldarelli}, \binits{G.}},
\bauthor{\bsnm{Quattrociocchi}, \binits{W.}}:
\batitle{Echo chambers: Emotional contagion and group polarization on facebook}.
\bjtitle{Scientific reports}
\bvolume{6}(\bissue{1}),
\bfpage{37825}
(\byear{2016})
\end{barticle}
\endbibitem

\bibitem[\protect\citeauthoryear{Del~Vicario et~al.}{2016b}]{del2016spreading}
\begin{barticle}
\bauthor{\bsnm{Del~Vicario}, \binits{M.}},
\bauthor{\bsnm{Bessi}, \binits{A.}},
\bauthor{\bsnm{Zollo}, \binits{F.}},
\bauthor{\bsnm{Petroni}, \binits{F.}},
\bauthor{\bsnm{Scala}, \binits{A.}},
\bauthor{\bsnm{Caldarelli}, \binits{G.}},
\bauthor{\bsnm{Stanley}, \binits{H.E.}},
\bauthor{\bsnm{Quattrociocchi}, \binits{W.}}:
\batitle{The spreading of misinformation online}.
\bjtitle{Proceedings of the national academy of Sciences}
\bvolume{113}(\bissue{3}),
\bfpage{554}--\blpage{559}
(\byear{2016})
\end{barticle}
\endbibitem

\bibitem[\protect\citeauthoryear{Choi et~al.}{2020}]{choi2020rumor}
\begin{barticle}
\bauthor{\bsnm{Choi}, \binits{D.}},
\bauthor{\bsnm{Chun}, \binits{S.}},
\bauthor{\bsnm{Oh}, \binits{H.}},
\bauthor{\bsnm{Han}, \binits{J.}},
\bauthor{\bsnm{Kwon}, \binits{T.T.}}:
\batitle{Rumor propagation is amplified by echo chambers in social media}.
\bjtitle{Scientific reports}
\bvolume{10}(\bissue{1}),
\bfpage{310}
(\byear{2020})
\end{barticle}
\endbibitem

\bibitem[\protect\citeauthoryear{Nyhan et~al.}{2023}]{nyhan2023like}
\begin{barticle}
\bauthor{\bsnm{Nyhan}, \binits{B.}},
\bauthor{\bsnm{Settle}, \binits{J.}},
\bauthor{\bsnm{Thorson}, \binits{E.}},
\bauthor{\bsnm{Wojcieszak}, \binits{M.}},
\bauthor{\bsnm{Barber{\'a}}, \binits{P.}},
\bauthor{\bsnm{Chen}, \binits{A.Y.}},
\bauthor{\bsnm{Allcott}, \binits{H.}},
\bauthor{\bsnm{Brown}, \binits{T.}},
\bauthor{\bsnm{Crespo-Tenorio}, \binits{A.}},
\bauthor{\bsnm{Dimmery}, \binits{D.}}, \betal:
\batitle{Like-minded sources on facebook are prevalent but not polarizing}.
\bjtitle{Nature}
\bvolume{620}(\bissue{7972}),
\bfpage{137}--\blpage{144}
(\byear{2023})
\end{barticle}
\endbibitem

\bibitem[\protect\citeauthoryear{Cinelli et~al.}{2021}]{cinelli2021echo}
\begin{barticle}
\bauthor{\bsnm{Cinelli}, \binits{M.}},
\bauthor{\bsnm{De~Francisci~Morales}, \binits{G.}},
\bauthor{\bsnm{Galeazzi}, \binits{A.}},
\bauthor{\bsnm{Quattrociocchi}, \binits{W.}},
\bauthor{\bsnm{Starnini}, \binits{M.}}:
\batitle{The echo chamber effect on social media}.
\bjtitle{Proceedings of the National Academy of Sciences}
\bvolume{118}(\bissue{9}),
\bfpage{2023301118}
(\byear{2021})
\end{barticle}
\endbibitem

\bibitem[\protect\citeauthoryear{Briand et~al.}{2021}]{briand2021infodemics}
\begin{barticle}
\bauthor{\bsnm{Briand}, \binits{S.C.}},
\bauthor{\bsnm{Cinelli}, \binits{M.}},
\bauthor{\bsnm{Nguyen}, \binits{T.}},
\bauthor{\bsnm{Lewis}, \binits{R.}},
\bauthor{\bsnm{Prybylski}, \binits{D.}},
\bauthor{\bsnm{Valensise}, \binits{C.M.}},
\bauthor{\bsnm{Colizza}, \binits{V.}},
\bauthor{\bsnm{Tozzi}, \binits{A.E.}},
\bauthor{\bsnm{Perra}, \binits{N.}},
\bauthor{\bsnm{Baronchelli}, \binits{A.}}, \betal:
\batitle{Infodemics: A new challenge for public health}.
\bjtitle{Cell}
\bvolume{184}(\bissue{25}),
\bfpage{6010}--\blpage{6014}
(\byear{2021})
\end{barticle}
\endbibitem

\bibitem[\protect\citeauthoryear{Perra and Rocha}{2019}]{perra2019modelling}
\begin{barticle}
\bauthor{\bsnm{Perra}, \binits{N.}},
\bauthor{\bsnm{Rocha}, \binits{L.E.}}:
\batitle{Modelling opinion dynamics in the age of algorithmic personalisation}.
\bjtitle{Scientific reports}
\bvolume{9}(\bissue{1}),
\bfpage{7261}
(\byear{2019})
\end{barticle}
\endbibitem

\bibitem[\protect\citeauthoryear{Guess et~al.}{2023}]{guess2023social}
\begin{barticle}
\bauthor{\bsnm{Guess}, \binits{A.M.}},
\bauthor{\bsnm{Malhotra}, \binits{N.}},
\bauthor{\bsnm{Pan}, \binits{J.}},
\bauthor{\bsnm{Barber{\'a}}, \binits{P.}},
\bauthor{\bsnm{Allcott}, \binits{H.}},
\bauthor{\bsnm{Brown}, \binits{T.}},
\bauthor{\bsnm{Crespo-Tenorio}, \binits{A.}},
\bauthor{\bsnm{Dimmery}, \binits{D.}},
\bauthor{\bsnm{Freelon}, \binits{D.}},
\bauthor{\bsnm{Gentzkow}, \binits{M.}}, \betal:
\batitle{How do social media feed algorithms affect attitudes and behavior in an election campaign?}
\bjtitle{Science}
\bvolume{381}(\bissue{6656}),
\bfpage{398}--\blpage{404}
(\byear{2023})
\end{barticle}
\endbibitem

\bibitem[\protect\citeauthoryear{Valensise et~al.}{2023}]{valensise2023drivers}
\begin{barticle}
\bauthor{\bsnm{Valensise}, \binits{C.M.}},
\bauthor{\bsnm{Cinelli}, \binits{M.}},
\bauthor{\bsnm{Quattrociocchi}, \binits{W.}}:
\batitle{The drivers of online polarization: Fitting models to data}.
\bjtitle{Information Sciences}
\bvolume{642},
\bfpage{119152}
(\byear{2023})
\end{barticle}
\endbibitem

\bibitem[\protect\citeauthoryear{Gonz{\'a}lez-Bail{\'o}n and Lelkes}{2023}]{gonzalez2023social}
\begin{barticle}
\bauthor{\bsnm{Gonz{\'a}lez-Bail{\'o}n}, \binits{S.}},
\bauthor{\bsnm{Lelkes}, \binits{Y.}}:
\batitle{Do social media undermine social cohesion? a critical review}.
\bjtitle{Social Issues and Policy Review}
\bvolume{17}(\bissue{1}),
\bfpage{155}--\blpage{180}
(\byear{2023})
\end{barticle}
\endbibitem

\bibitem[\protect\citeauthoryear{Gonz{\'a}lez-Bail{\'o}n et~al.}{2023}]{gonzalez2023asymmetric}
\begin{barticle}
\bauthor{\bsnm{Gonz{\'a}lez-Bail{\'o}n}, \binits{S.}},
\bauthor{\bsnm{Lazer}, \binits{D.}},
\bauthor{\bsnm{Barber{\'a}}, \binits{P.}},
\bauthor{\bsnm{Zhang}, \binits{M.}},
\bauthor{\bsnm{Allcott}, \binits{H.}},
\bauthor{\bsnm{Brown}, \binits{T.}},
\bauthor{\bsnm{Crespo-Tenorio}, \binits{A.}},
\bauthor{\bsnm{Freelon}, \binits{D.}},
\bauthor{\bsnm{Gentzkow}, \binits{M.}},
\bauthor{\bsnm{Guess}, \binits{A.M.}}, \betal:
\batitle{Asymmetric ideological segregation in exposure to political news on facebook}.
\bjtitle{Science}
\bvolume{381}(\bissue{6656}),
\bfpage{392}--\blpage{398}
(\byear{2023})
\end{barticle}
\endbibitem

\bibitem[\protect\citeauthoryear{Simon and Greenberger}{1971}]{simon1971computers}
\begin{botherref}
\oauthor{\bsnm{Simon}, \binits{H.}},
\oauthor{\bsnm{Greenberger}, \binits{M.}}:
Computers, communications and the public interest.
Computers, communications, and the public interest. Johns Hopkins Press, Baltimore,
40--41
(1971)
\end{botherref}
\endbibitem

\bibitem[\protect\citeauthoryear{Davenport and Beck}{2001}]{davenport2001attention}
\begin{barticle}
\bauthor{\bsnm{Davenport}, \binits{T.H.}},
\bauthor{\bsnm{Beck}, \binits{J.C.}}:
\batitle{The attention economy}.
\bjtitle{Ubiquity}
\bvolume{2001}(\bissue{May}),
\bfpage{1}
(\byear{2001})
\end{barticle}
\endbibitem

\bibitem[\protect\citeauthoryear{Falkinger}{2007}]{falkinger2007attention}
\begin{barticle}
\bauthor{\bsnm{Falkinger}, \binits{J.}}:
\batitle{Attention economies}.
\bjtitle{Journal of Economic Theory}
\bvolume{133}(\bissue{1}),
\bfpage{266}--\blpage{294}
(\byear{2007})
\end{barticle}
\endbibitem

\bibitem[\protect\citeauthoryear{Falkinger}{2008}]{falkinger2008limited}
\begin{barticle}
\bauthor{\bsnm{Falkinger}, \binits{J.}}:
\batitle{Limited attention as a scarce resource in information-rich economies}.
\bjtitle{The Economic Journal}
\bvolume{118}(\bissue{532}),
\bfpage{1596}--\blpage{1620}
(\byear{2008})
\end{barticle}
\endbibitem

\bibitem[\protect\citeauthoryear{Anderson and De~Palma}{2012}]{anderson2012competition}
\begin{barticle}
\bauthor{\bsnm{Anderson}, \binits{S.P.}},
\bauthor{\bsnm{De~Palma}, \binits{A.}}:
\batitle{Competition for attention in the information (overload) age}.
\bjtitle{The RAND Journal of Economics}
\bvolume{43}(\bissue{1}),
\bfpage{1}--\blpage{25}
(\byear{2012})
\end{barticle}
\endbibitem

\bibitem[\protect\citeauthoryear{Weng et~al.}{2012}]{weng2012competition}
\begin{barticle}
\bauthor{\bsnm{Weng}, \binits{L.}},
\bauthor{\bsnm{Flammini}, \binits{A.}},
\bauthor{\bsnm{Vespignani}, \binits{A.}},
\bauthor{\bsnm{Menczer}, \binits{F.}}:
\batitle{Competition among memes in a world with limited attention}.
\bjtitle{Scientific reports}
\bvolume{2}(\bissue{1}),
\bfpage{335}
(\byear{2012})
\end{barticle}
\endbibitem

\bibitem[\protect\citeauthoryear{Tufekci}{2013}]{tufekci2013not}
\begin{barticle}
\bauthor{\bsnm{Tufekci}, \binits{Z.}}:
\batitle{“not this one” social movements, the attention economy, and microcelebrity networked activism}.
\bjtitle{American behavioral scientist}
\bvolume{57}(\bissue{7}),
\bfpage{848}--\blpage{870}
(\byear{2013})
\end{barticle}
\endbibitem

\bibitem[\protect\citeauthoryear{Lorenz-Spreen et~al.}{2019}]{lorenz2019accelerating}
\begin{barticle}
\bauthor{\bsnm{Lorenz-Spreen}, \binits{P.}},
\bauthor{\bsnm{M{\o}nsted}, \binits{B.M.}},
\bauthor{\bsnm{H{\"o}vel}, \binits{P.}},
\bauthor{\bsnm{Lehmann}, \binits{S.}}:
\batitle{Accelerating dynamics of collective attention}.
\bjtitle{Nature communications}
\bvolume{10}(\bissue{1}),
\bfpage{1759}
(\byear{2019})
\end{barticle}
\endbibitem

\bibitem[\protect\citeauthoryear{Bhargava and Velasquez}{2021}]{bhargava2021ethics}
\begin{barticle}
\bauthor{\bsnm{Bhargava}, \binits{V.R.}},
\bauthor{\bsnm{Velasquez}, \binits{M.}}:
\batitle{Ethics of the attention economy: The problem of social media addiction}.
\bjtitle{Business Ethics Quarterly}
\bvolume{31}(\bissue{3}),
\bfpage{321}--\blpage{359}
(\byear{2021})
\end{barticle}
\endbibitem

\bibitem[\protect\citeauthoryear{Ciampaglia et~al.}{2015}]{ciampaglia2015production}
\begin{barticle}
\bauthor{\bsnm{Ciampaglia}, \binits{G.L.}},
\bauthor{\bsnm{Flammini}, \binits{A.}},
\bauthor{\bsnm{Menczer}, \binits{F.}}:
\batitle{The production of information in the attention economy}.
\bjtitle{Scientific reports}
\bvolume{5}(\bissue{1}),
\bfpage{9452}
(\byear{2015})
\end{barticle}
\endbibitem

\bibitem[\protect\citeauthoryear{Gibrat}{1931}]{gibrat1931inegalits}
\begin{botherref}
\oauthor{\bsnm{Gibrat}, \binits{R.}}:
Les in{\'e}galits {\'e}conomiques.
Sirey
(1931)
\end{botherref}
\endbibitem

\bibitem[\protect\citeauthoryear{Rozenfeld et~al.}{2008}]{rozenfeld2008laws}
\begin{barticle}
\bauthor{\bsnm{Rozenfeld}, \binits{H.D.}},
\bauthor{\bsnm{Rybski}, \binits{D.}},
\bauthor{\bsnm{Andrade~Jr}, \binits{J.S.}},
\bauthor{\bsnm{Batty}, \binits{M.}},
\bauthor{\bsnm{Stanley}, \binits{H.E.}},
\bauthor{\bsnm{Makse}, \binits{H.A.}}:
\batitle{Laws of population growth}.
\bjtitle{Proceedings of the National Academy of Sciences}
\bvolume{105}(\bissue{48}),
\bfpage{18702}--\blpage{18707}
(\byear{2008})
\end{barticle}
\endbibitem

\bibitem[\protect\citeauthoryear{Eeckhout}{2004}]{eeckhout2004gibrat}
\begin{barticle}
\bauthor{\bsnm{Eeckhout}, \binits{J.}}:
\batitle{Gibrat's law for (all) cities}.
\bjtitle{American Economic Review}
\bvolume{94}(\bissue{5}),
\bfpage{1429}--\blpage{1451}
(\byear{2004})
\end{barticle}
\endbibitem

\bibitem[\protect\citeauthoryear{Rose}{2005}]{rose2005cities}
\begin{botherref}
\oauthor{\bsnm{Rose}, \binits{A.K.}}:
Cities and countries.
National Bureau of Economic Research Cambridge, Mass., USA
(2005)
\end{botherref}
\endbibitem

\bibitem[\protect\citeauthoryear{Mansfield}{1962}]{mansfield1962entry}
\begin{barticle}
\bauthor{\bsnm{Mansfield}, \binits{E.}}:
\batitle{Entry, gibrat's law, innovation, and the growth of firms}.
\bjtitle{The American economic review}
\bvolume{52}(\bissue{5}),
\bfpage{1023}--\blpage{1051}
(\byear{1962})
\end{barticle}
\endbibitem

\bibitem[\protect\citeauthoryear{Chesher}{1979}]{chesher1979testing}
\begin{barticle}
\bauthor{\bsnm{Chesher}, \binits{A.}}:
\batitle{Testing the law of proportionate effect}.
\bjtitle{The Journal of industrial economics}
\bvolume{27}(\bissue{4}),
\bfpage{403}--\blpage{411}
(\byear{1979})
\end{barticle}
\endbibitem

\bibitem[\protect\citeauthoryear{Sutton}{1997}]{sutton1997gibrat}
\begin{barticle}
\bauthor{\bsnm{Sutton}, \binits{J.}}:
\batitle{Gibrat's legacy}.
\bjtitle{Journal of economic literature}
\bvolume{35}(\bissue{1}),
\bfpage{40}--\blpage{59}
(\byear{1997})
\end{barticle}
\endbibitem

\bibitem[\protect\citeauthoryear{Santarelli et~al.}{2006}]{santarelli2006gibrat}
\begin{botherref}
\oauthor{\bsnm{Santarelli}, \binits{E.}},
\oauthor{\bsnm{Klomp}, \binits{L.}},
\oauthor{\bsnm{Thurik}, \binits{A.R.}}:
Gibrat’s law: An overview of the empirical literature.
Entrepreneurship, growth, and innovation: The dynamics of firms and industries,
41--73
(2006)
\end{botherref}
\endbibitem

\bibitem[\protect\citeauthoryear{}{2023}]{newsguard}
\begin{botherref}
NewsGuard.
https://www.newsguardtech.com
(2023)
\end{botherref}
\endbibitem

\bibitem[\protect\citeauthoryear{}{2023}]{crowdtangle}
\begin{botherref}
CrowdTangle.
CrowdTangle Team. Facebook, Menlo Park, California, United States
(2023)
\end{botherref}
\endbibitem

\bibitem[\protect\citeauthoryear{Plerou et~al.}{1999}]{plerou1999similarities}
\begin{barticle}
\bauthor{\bsnm{Plerou}, \binits{V.}},
\bauthor{\bsnm{Amaral}, \binits{L.A.N.}},
\bauthor{\bsnm{Gopikrishnan}, \binits{P.}},
\bauthor{\bsnm{Meyer}, \binits{M.}},
\bauthor{\bsnm{Stanley}, \binits{H.E.}}:
\batitle{Similarities between the growth dynamics of university research and of competitive economic activities}.
\bjtitle{Nature}
\bvolume{400}(\bissue{6743}),
\bfpage{433}--\blpage{437}
(\byear{1999})
\end{barticle}
\endbibitem

\bibitem[\protect\citeauthoryear{Qian et~al.}{2014}]{qian2014origin}
\begin{barticle}
\bauthor{\bsnm{Qian}, \binits{J.-H.}},
\bauthor{\bsnm{Chen}, \binits{Q.}},
\bauthor{\bsnm{Han}, \binits{D.-D.}},
\bauthor{\bsnm{Ma}, \binits{Y.-G.}},
\bauthor{\bsnm{Shen}, \binits{W.-Q.}}:
\batitle{Origin of gibrat law in internet: Asymmetric distribution of the correlation}.
\bjtitle{Physical Review E}
\bvolume{89}(\bissue{6}),
\bfpage{062808}
(\byear{2014})
\end{barticle}
\endbibitem

\bibitem[\protect\citeauthoryear{Stanley et~al.}{1996}]{stanley1996scaling}
\begin{barticle}
\bauthor{\bsnm{Stanley}, \binits{M.H.}},
\bauthor{\bsnm{Amaral}, \binits{L.A.}},
\bauthor{\bsnm{Buldyrev}, \binits{S.V.}},
\bauthor{\bsnm{Havlin}, \binits{S.}},
\bauthor{\bsnm{Leschhorn}, \binits{H.}},
\bauthor{\bsnm{Maass}, \binits{P.}},
\bauthor{\bsnm{Salinger}, \binits{M.A.}},
\bauthor{\bsnm{Stanley}, \binits{H.E.}}:
\batitle{Scaling behaviour in the growth of companies}.
\bjtitle{Nature}
\bvolume{379}(\bissue{6568}),
\bfpage{804}--\blpage{806}
(\byear{1996})
\end{barticle}
\endbibitem

\bibitem[\protect\citeauthoryear{Amaral et~al.}{1997}]{amaral1997scaling}
\begin{barticle}
\bauthor{\bsnm{Amaral}, \binits{L.A.N.}},
\bauthor{\bsnm{Buldyrev}, \binits{S.V.}},
\bauthor{\bsnm{Havlin}, \binits{S.}},
\bauthor{\bsnm{Leschhorn}, \binits{H.}},
\bauthor{\bsnm{Maass}, \binits{P.}},
\bauthor{\bsnm{Salinger}, \binits{M.A.}},
\bauthor{\bsnm{Stanley}, \binits{H.E.}},
\bauthor{\bsnm{Stanley}, \binits{M.H.}}:
\batitle{Scaling behavior in economics: I. empirical results for company growth}.
\bjtitle{Journal de Physique I}
\bvolume{7}(\bissue{4}),
\bfpage{621}--\blpage{633}
(\byear{1997})
\end{barticle}
\endbibitem

\bibitem[\protect\citeauthoryear{Burr}{1942}]{burr1942cumulative}
\begin{barticle}
\bauthor{\bsnm{Burr}, \binits{I.W.}}:
\batitle{Cumulative frequency functions}.
\bjtitle{The Annals of mathematical statistics}
\bvolume{13}(\bissue{2}),
\bfpage{215}--\blpage{232}
(\byear{1942})
\end{barticle}
\endbibitem

\bibitem[\protect\citeauthoryear{Fujiwara et~al.}{2003}]{fujiwara2003growth}
\begin{barticle}
\bauthor{\bsnm{Fujiwara}, \binits{Y.}},
\bauthor{\bsnm{Souma}, \binits{W.}},
\bauthor{\bsnm{Aoyama}, \binits{H.}},
\bauthor{\bsnm{Kaizoji}, \binits{T.}},
\bauthor{\bsnm{Aoki}, \binits{M.}}:
\batitle{Growth and fluctuations of personal income}.
\bjtitle{Physica A: Statistical Mechanics and its Applications}
\bvolume{321}(\bissue{3-4}),
\bfpage{598}--\blpage{604}
(\byear{2003})
\end{barticle}
\endbibitem

\bibitem[\protect\citeauthoryear{Fujiwara et~al.}{2004}]{fujiwara2004pareto}
\begin{barticle}
\bauthor{\bsnm{Fujiwara}, \binits{Y.}},
\bauthor{\bsnm{Di~Guilmi}, \binits{C.}},
\bauthor{\bsnm{Aoyama}, \binits{H.}},
\bauthor{\bsnm{Gallegati}, \binits{M.}},
\bauthor{\bsnm{Souma}, \binits{W.}}:
\batitle{Do pareto--zipf and gibrat laws hold true? an analysis with european firms}.
\bjtitle{Physica A: Statistical Mechanics and its Applications}
\bvolume{335}(\bissue{1-2}),
\bfpage{197}--\blpage{216}
(\byear{2004})
\end{barticle}
\endbibitem

\bibitem[\protect\citeauthoryear{Etta et~al.}{2023}]{etta2023characterizing}
\begin{barticle}
\bauthor{\bsnm{Etta}, \binits{G.}},
\bauthor{\bsnm{Sangiorgio}, \binits{E.}},
\bauthor{\bsnm{Di~Marco}, \binits{N.}},
\bauthor{\bsnm{Avalle}, \binits{M.}},
\bauthor{\bsnm{Scala}, \binits{A.}},
\bauthor{\bsnm{Cinelli}, \binits{M.}},
\bauthor{\bsnm{Quattrociocchi}, \binits{W.}}:
\batitle{Characterizing engagement dynamics across topics on facebook}.
\bjtitle{Plos one}
\bvolume{18}(\bissue{6}),
\bfpage{0286150}
(\byear{2023})
\end{barticle}
\endbibitem

\bibitem[\protect\citeauthoryear{Bakshy et~al.}{2011}]{bakshy2011everyone}
\begin{bchapter}
\bauthor{\bsnm{Bakshy}, \binits{E.}},
\bauthor{\bsnm{Hofman}, \binits{J.M.}},
\bauthor{\bsnm{Mason}, \binits{W.A.}},
\bauthor{\bsnm{Watts}, \binits{D.J.}}:
\bctitle{Everyone's an influencer: quantifying influence on twitter}.
In: \bbtitle{Proceedings of the Fourth ACM International Conference on Web Search and Data Mining},
pp. \bfpage{65}--\blpage{74}
(\byear{2011})
\end{bchapter}
\endbibitem

\bibitem[\protect\citeauthoryear{Cresci}{2020}]{cresci2020decade}
\begin{barticle}
\bauthor{\bsnm{Cresci}, \binits{S.}}:
\batitle{A decade of social bot detection}.
\bjtitle{Communications of the ACM}
\bvolume{63}(\bissue{10}),
\bfpage{72}--\blpage{83}
(\byear{2020})
\end{barticle}
\endbibitem

\end{thebibliography}

\newpage

\section{Supporting Information}

\subsection{Data Collection}

We download our data from CrowdTangle, a Facebook-owned tool that monitors interactions on public content from Facebook pages, groups, and verified profiles. CrowdTangle is accessible to researchers upon request at this link \url{https://www.crowdtangle.com/request}. We obtain the list of news outlets employed in the analysis via NewsGuard, which provides, for each outlet, more than 30 distinct categories of descriptive metadata, including a breakdown of their assessment, an indication of its political slant, descriptions of the topics—or types of misinformation—it covers, and more. After selecting all the news outlets with a Facebook account listed on NewsGuard, we use their Facebook URLs to gather their data on CrowdTangle. Using the tool 'Historical Data' provided by CrowdTangle, we download the entire history of each page from its creation date as a table containing information regarding each posted item in chronological order.

\subsection{Processing Methods}
\begin{figure}
\centering
\includegraphics[scale=0.5]{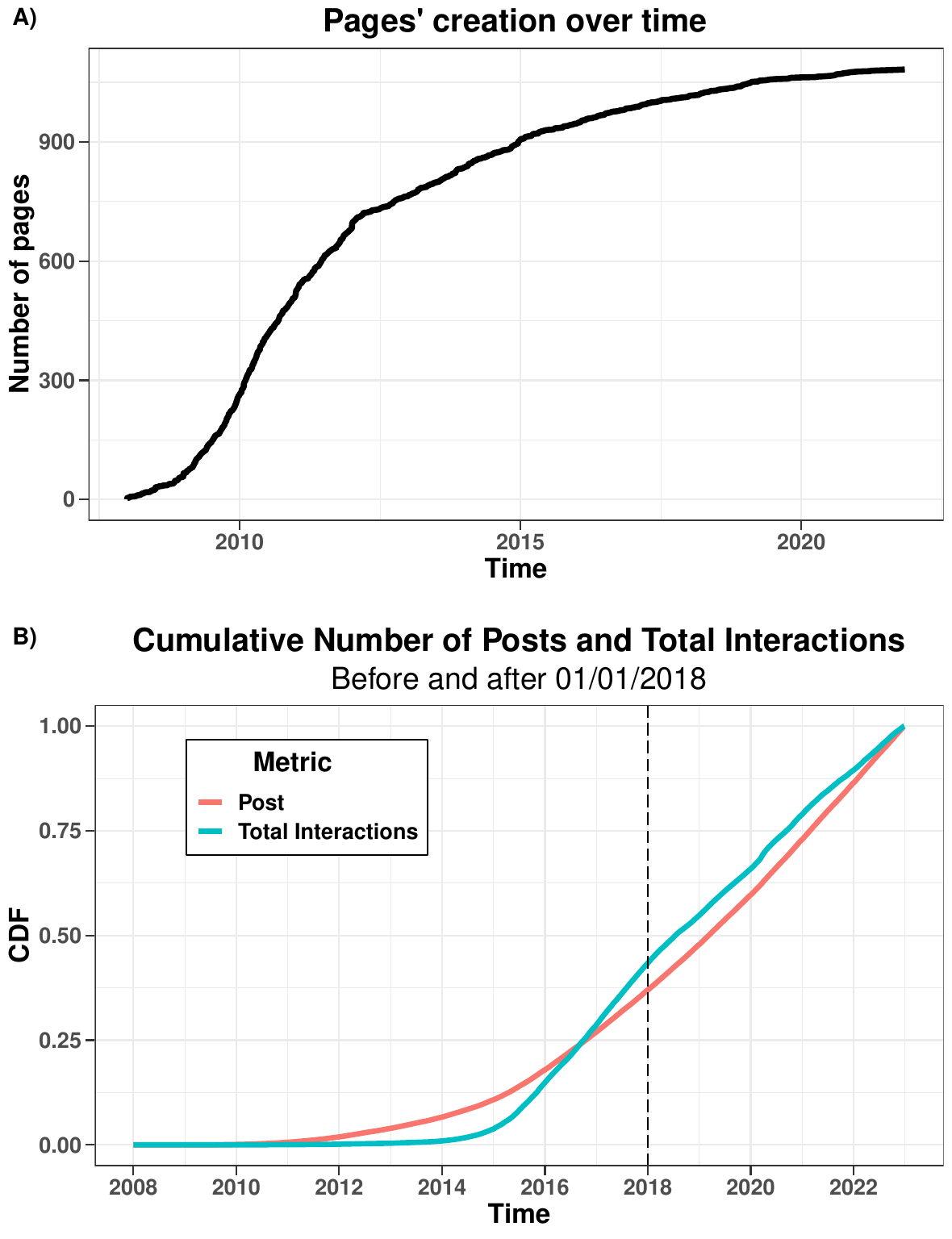}
\caption{(A) Pages' creation across time. (B) Evolution of the number of posts and Total Interactions over time.}
\label{fig:SI_pages}
\end{figure}

With a post-level granularity, the table's columns include several relevant fields, including the post type (link, image, video), its text, the number of reactions to the post, and more. Among these columns, we select two relevant ones: \textit{Total Interactions} and \textit{Followers at Posting}. The Total Interactions column contains the total number of reactions per post (that is, the sum of Likes, Comments, and Shares) that we aggregate (sum) depending on the time scale of the analysis. The Total Interactions are thus our metric accounting for the Engagement. These data are available for the whole page history, that is, for the longest-running pages, from the beginning of 2008 to the end of 2022, when we downloaded data. Since pages' creation dates span through time, our analysis of growth rates is independent of the creation date once accounted for a sufficient life span and activity of the page. The cumulative time series regarding existing pages starting to be active on Facebook is reported in panel A of Fig. \ref{fig:SI_pages}. The number of pages included in the dataset is 1082, and 94 \% of them were created before 01/01/2018. The second variable we use for our analysis is Followers at Posting, which represents the number of users subscribed to a given page at the time of posting. This information, however, is only available for posts made since 01/01/2018. Before that day, CrowdTangle was not collecting such information, or it is not sharing it with end users as of now. For this reason, when we consider the metrics of Followers and Engagement jointly, we restrict our analysis period to 01/01/2018 - 31/12/2022. In the analysis in which we do not account for Followers' value, we consider the entire 15-year timespan. Quantitatively speaking, in panel B of Fig. \ref{fig:SI_pages}, we report the evolution of the number of posts we consider in the analysis, which is around 57 million, and the Total Interactions, which is around 21 billion, over time. We note from 1/1/2018 onward, 63 \% of posts and 56 \% of Total Interactions were produced. 

\subsection{Test Results}
Table S1 shows p-values of Mann-Whitney U tests between Questionable and Reliable pages. The alternative hypothesis is that the distribution of Reliable sources is right-shifted (higher mean) with respect to the Questionable one. Engagement indicate the absolute value distributions, Engagement growth indicate the growth rate ones. This comparison based on information quality is consistent with overall evidence. The engagement growth of unreliable pages progressively became negative as the time scale widens, in Monthly and Quarterly scales, despite having constantly lower absolute Engagement. Anew, the short-term fluctuations follow a universal dynamic, and neither the reliability turns out to determine growth differences.

\begin{table}[ht]
\centering
\begin{tabular}{lcl}
  \hline
Timescale & p-value & Metric \\ 
  \hline
Day & 0.00 & Engagement \\ 
Week & 0.00 & Engagement \\ 
Month & 0.00 & Engagement \\ 
Quarter & 0.00 & Engagement \\ 
Day & 0.32 & Engagement growth \\ 
Week & 0.10 & Engagement growth \\ 
Month & 0.00 & Engagement growth \\ 
Quarter & 0.00 & Engagement growth \\ 
\hline
\end{tabular}
\label{tab:SI_reliability_test}
\caption{p-values of Mann-Whitney U tests between Questionable and Reliable pages. H1: The distribution of Reliable sources is right-shifted (higher mean) with respect to the Questionable one.}
\end{table}

\newpage
\begin{figure}
\centering
\includegraphics[scale=0.5]{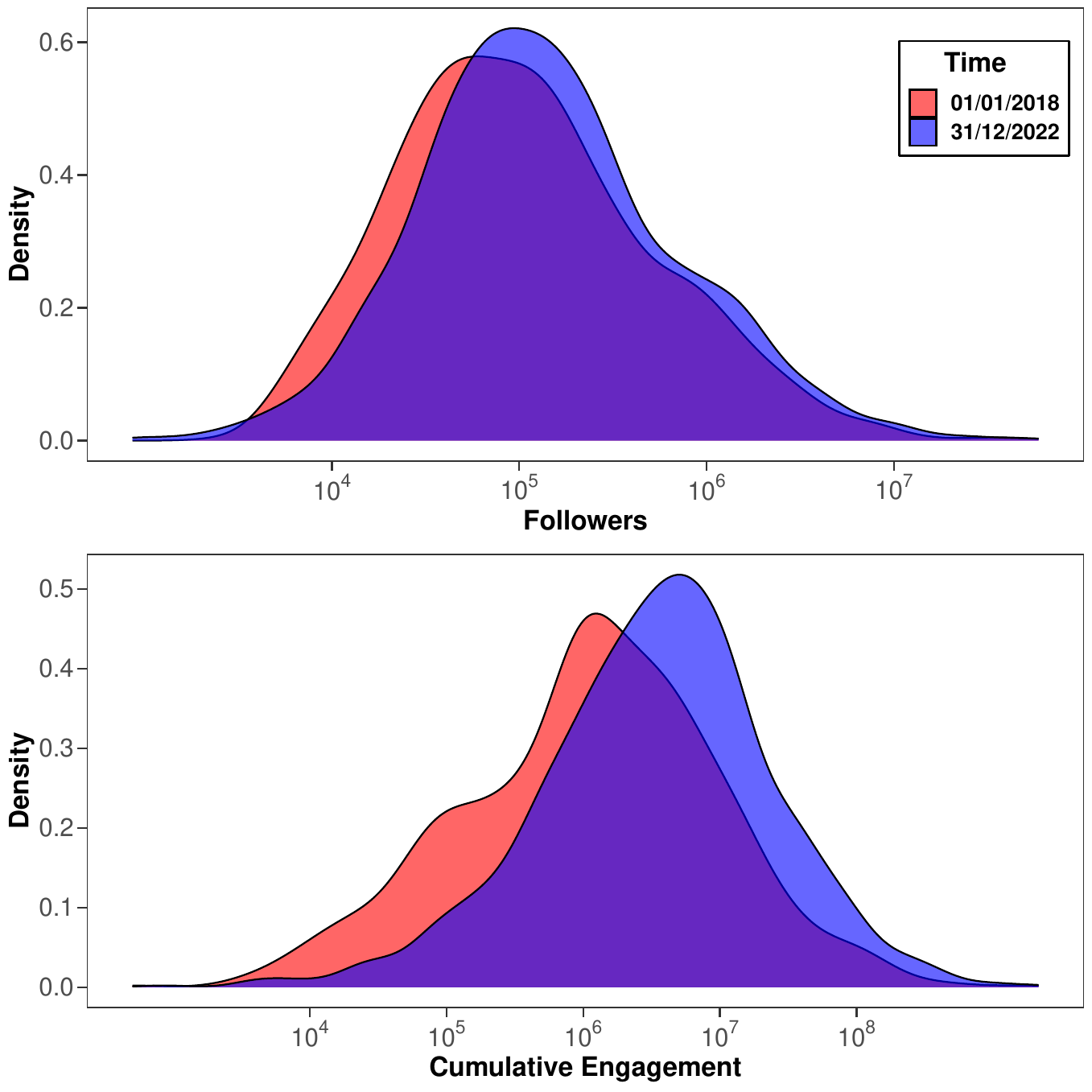}
\caption{Distributions of both possible size indicators, Followers and Cumulative Engagement, of the entire sample at the start and end of our analysis period. Both distributions manifest as heavy-tailed, here displayed on a logarithmic scale.}
\label{fig:SI_lognormal}
\end{figure}

\begin{figure}
\centering
\includegraphics[scale=0.5]{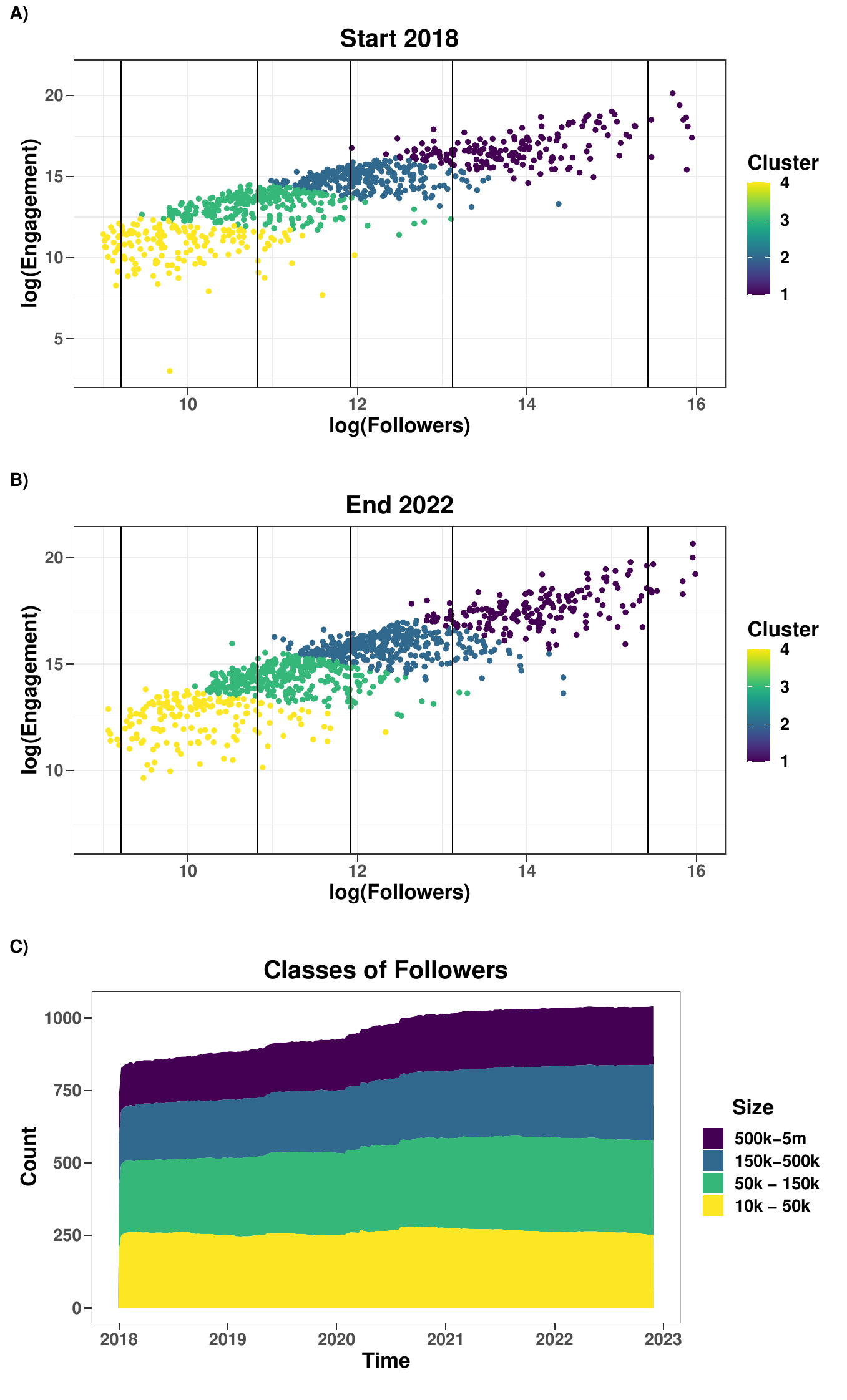}
\caption{(A-B) Partitions of PAM clustering of news outlets, based on Followers and Cumulative Engagement values at the start (1/1/2018) and the end (31/12/2022) of our analysis period, respectively. Solid vertical lines represent the range limits of our selected classes of Followers. (C) Frequencies of the size classes across the analysis period.}
\label{fig:SI_size_frequencies}
\end{figure}

\newpage
\begin{figure}
\centering
\includegraphics[width=\textwidth]{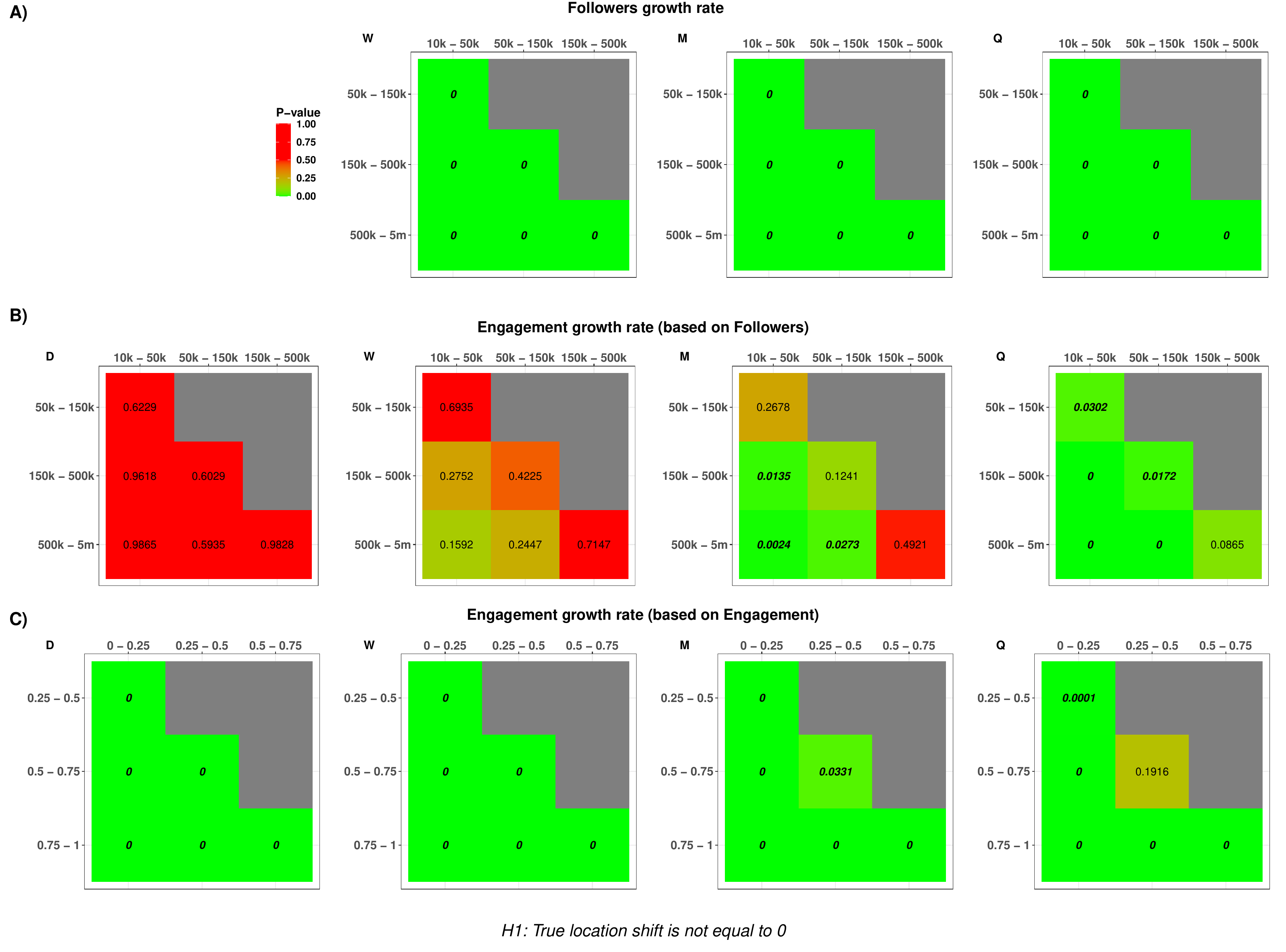}
\caption{(A-C) p-values of two-sided Mann-Whitney U tests between classes of size for Followers and Engagement growth rate distributions. Panel titles indicate the metric being tested and the metric according to which we determine the size. Row and column headers represent the class size. Bold numbers represent p-values for which we reject the hypothesis that the growth distributions do not differ, with the alternative hypothesis that the true location shift is not equal to 0. For readability, 0 represents p-values smaller than 0.0001.}
\label{fig:SI_twoside}
\end{figure}

\newpage
\begin{figure}
\centering
\includegraphics[width=\textwidth]{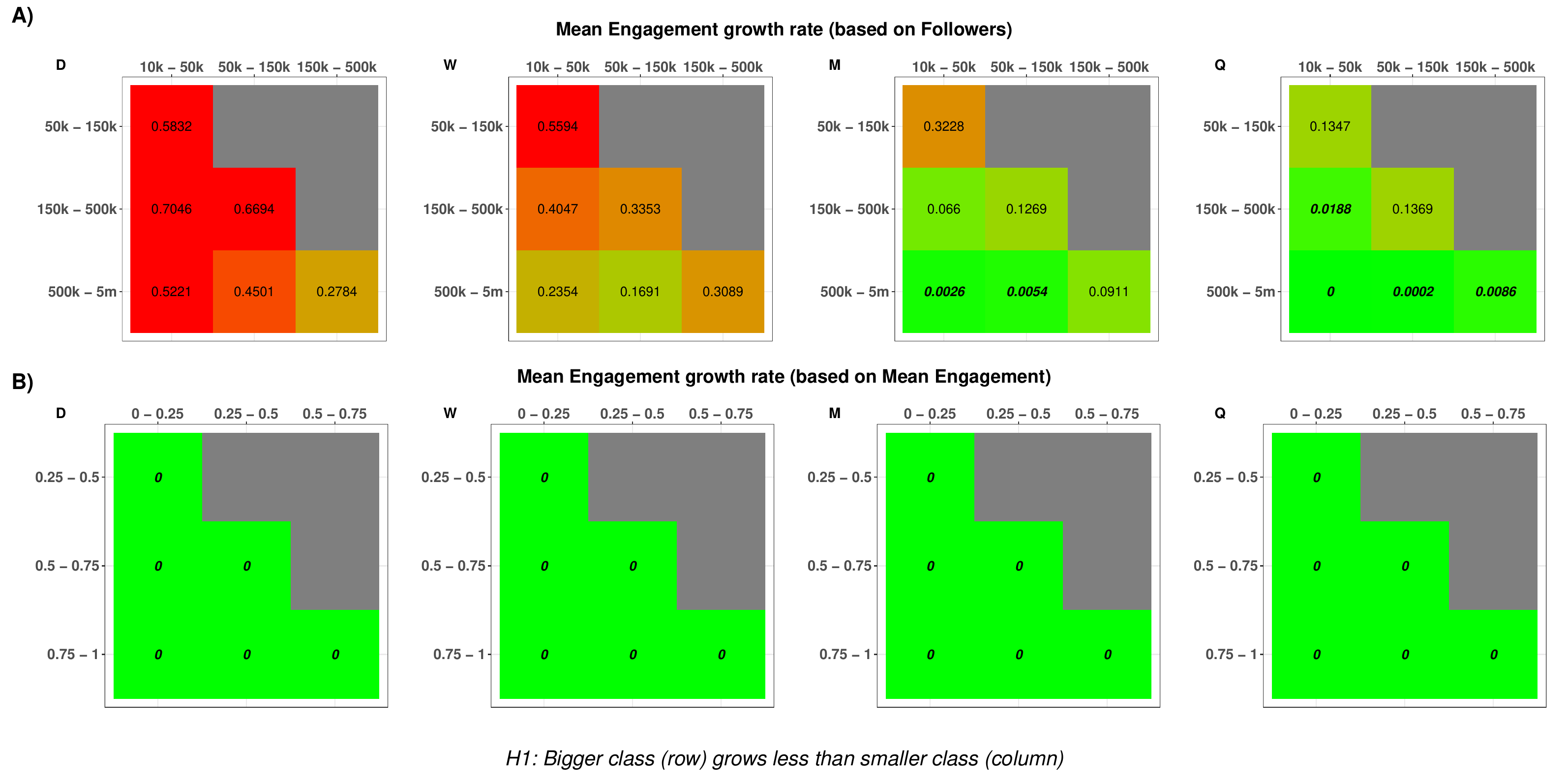}
\caption{We recalculated the Engagement using its mean value and repeated the tests reported in Fig. 1B and 1C. (A-C) p-values of two-sided Mann-Whitney U tests between classes of size for Followers and Engagement growth rate distributions. Panel titles indicate the metric being tested and the metric according to which we determine the size. Row and column headers represent the class size. Bold numbers represent p-values for which we reject the hypothesis that the growth distributions do not differ, with the alternative hypothesis that the smaller class grows at a higher rate. For readability, 0 represents p-values smaller than 0.0001. }
\label{fig:SI_mean_eng}
\end{figure}

\newpage
\begin{figure}
\centering
\includegraphics[width=\textwidth]{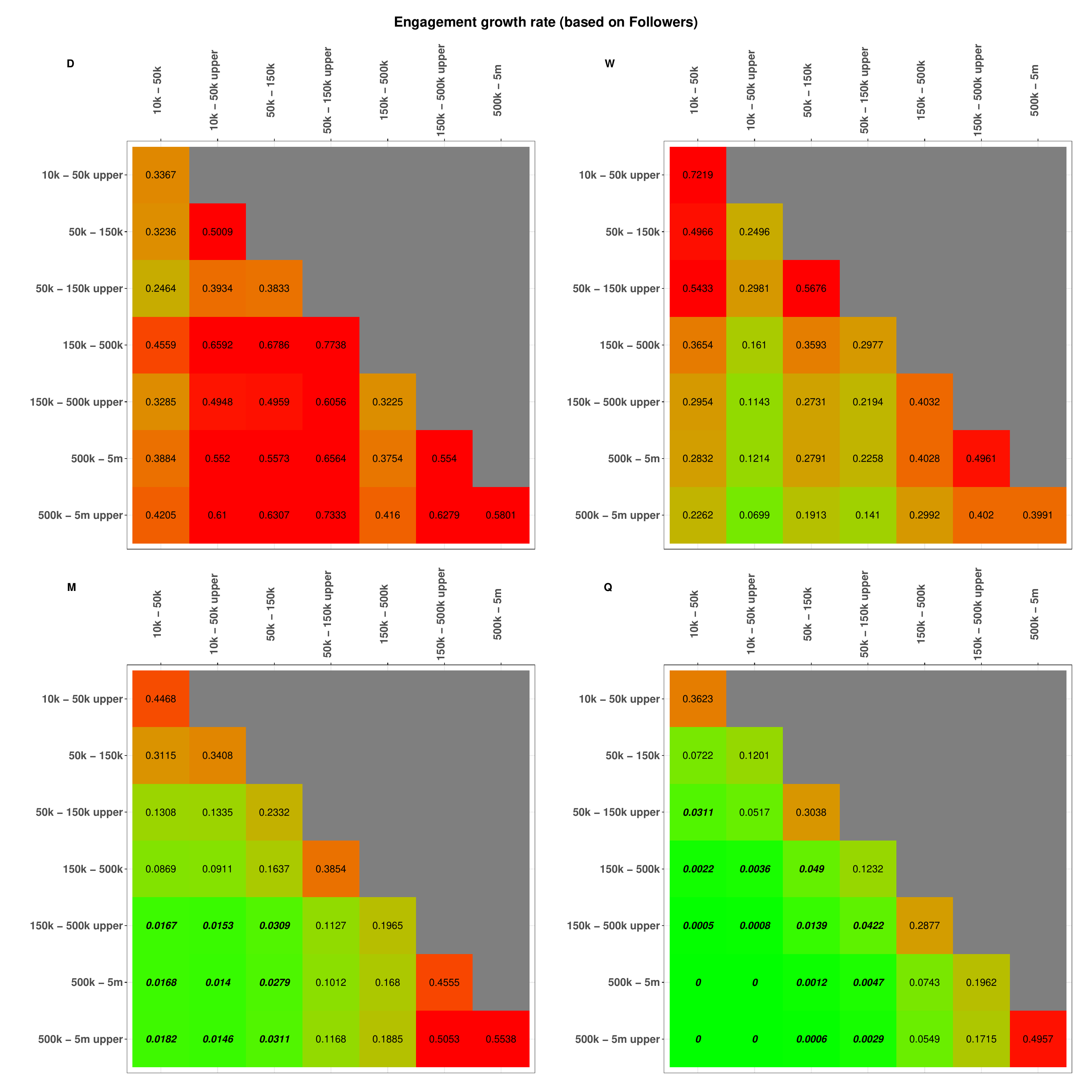}
\caption{As the growth dynamic emerges clearly across the different timescales, changing the bin boundaries should not lead to changes in the results.
We repeated the tests reported in Fig. 1B and 1C, dividing each original bin into two sub-classes of an equal number of observations by cutting the bin through its Median Followers Value. The independence of growth from size in the short term still holds, with none of the 28 tests showing differences in growth in both Daily and Weekly measurements. On the monthly scale, the three smaller classes grow faster than the three bigger ones, while on the quarterly scale, most tests show statistically significant p-values, except for the pairs of adjacent classes.
(A-C) p-values of two-sided Mann-Whitney U tests between classes of size for Followers and Engagement growth rate distributions. Panel titles indicate the metric being tested and the metric according to which we determine the size. Row and column headers represent the class size. Bold numbers represent p-values for which we reject the hypothesis that the growth distributions do not differ, with the alternative hypothesis that the smaller class grows at a higher rate. For readability, 0 represents p-values smaller than 0.0001. }
\label{fig:SI_8_bins}
\end{figure}

\newpage
\begin{figure}
\centering
\includegraphics[scale=0.5]{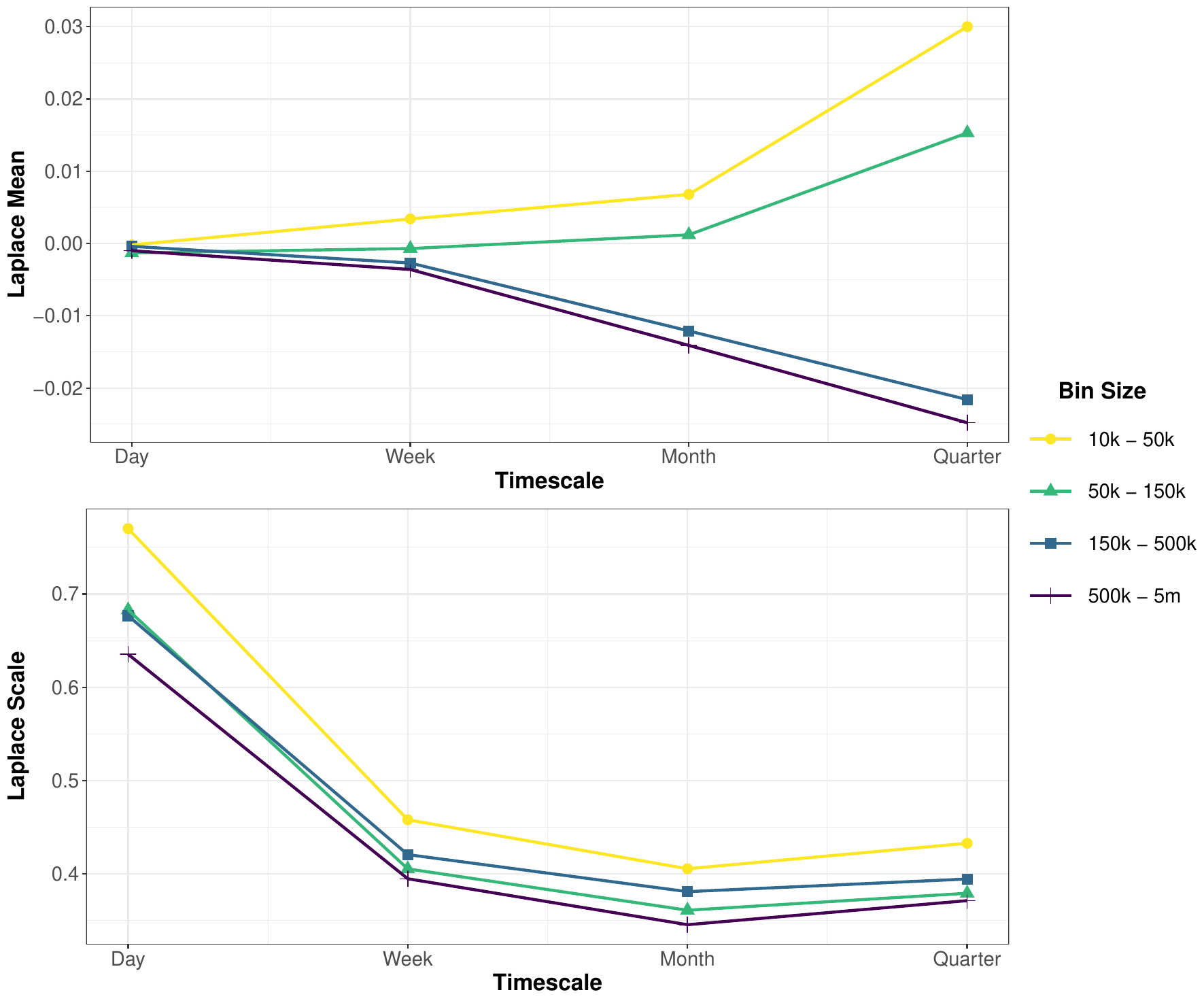}
\caption{Laplace parameter variation with the increase of the observed timescale for the four size classes. Upper panel represent $\mu$ parameter variation, lower panel represent $b$ parameter variation. As upper panel shows, $\mu \to 0$ with the narrowing of the observed timescale.}
\label{fig:SI_Laplace}
\end{figure}

\newpage
\begin{figure}
\centering
\includegraphics[scale=0.5]{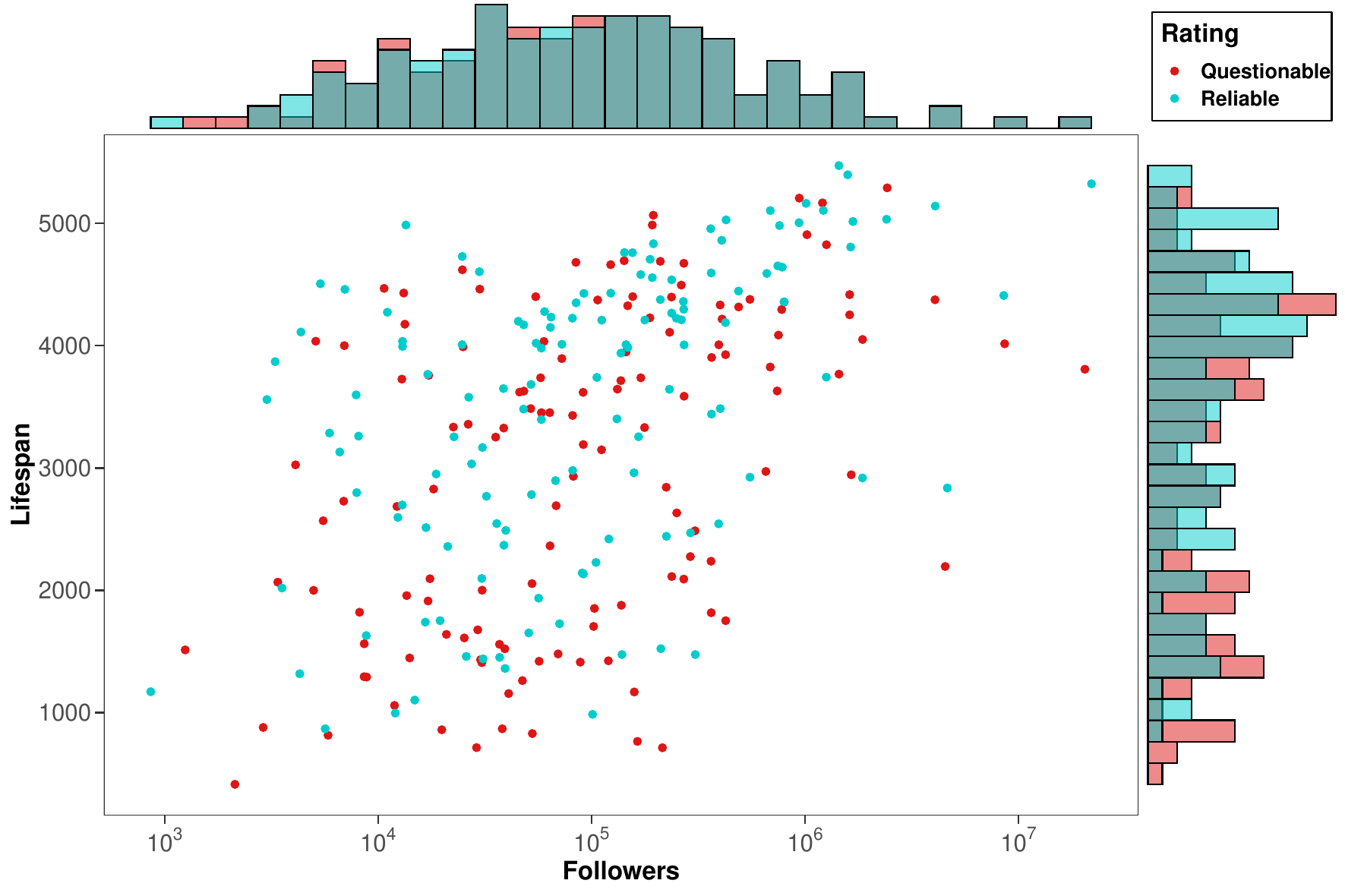}
\caption{Scatterplot of Questionable and Reliable pages sub-sample. According to NewsGuard ratings, our dataset comprises 898 reliable sources and 131 non-reliable ones. We performed a sampling of 131 reliable sources to have two comparable samples. The resulting sample is obtained by selecting the reliable pages for which the overall Euclidean distance from the non-reliable sample is minimized. We aim to achieve similar structural characteristics that are not being tested, namely Followers and Lifespan (here shown in days). Therefore, we compute distance using the number of Followers and the page's creation date as distance variables since most pages' last observations coincide with the end of the analyzed period.}
\label{fig:SI_rel_scatter}
\end{figure}

\end{document}